\documentclass[twocolumn,aps,prl,groupedaddress]{revtex4}
\usepackage{amssymb}
\usepackage{amsmath}
\usepackage{color}
\usepackage{graphicx}

\newcommand{\be}{\begin{equation}}
\newcommand{\ee}{\end{equation}}
\newcommand{\nn}{\nonumber\\}

\newcommand{\p}{\partial}
\newcommand{\la}{\langle}
\newcommand{\ra}{\rangle}
\newcommand{\lb}{\left[}
\newcommand{\rb}{\right]}
\newcommand{\lp}{\left(}
\newcommand{\rp}{\right)}

\renewcommand{\vec}[1]{{\bf #1}}

\begin{document}

\title{Gate-tunable flat bands in van der Waals patterned dielectric superlattices}
\author{Li-kun Shi$^1$, Jing Ma$^2$, and Justin C.W. Song$^{1,2}$}
\email{justinsong@ntu.edu.sg}
\affiliation{$^1$Institute of High Performance Computing, Agency for Science, Technology, \& Research, Singapore 138632}
\affiliation{$^2$Division of Physics and Applied Physics, Nanyang Technological University, Singapore 637371}

\begin{abstract}
Superlattice engineering provides the means to reshape the fabric felt by quasiparticles moving in a  material. Here we argue that bandstructure engineering with superlattices can be pushed to the extreme limit by stacking gapped van der Waals (vdW) materials on patterned dielectric substrates.
Specifically, we find that high quality vdW patterned dielectric superlattices (PDS) realize a series of robust flat bands that can be directly switched on and off by gate voltage in situ. 
In contrast to existing superlattice platforms, these flat bands are realized without the need for fine tuning. Instead, the bands become flat as the gate voltage increases in magnitude. 
The characteristics of PDS flatbands are highly tunable: the type of flatband (single non-degenerate or dirac-cone-like), localization length, and interaction energy are sensitive to the applied gate voltage. As a result, electron-electron interactions in the PDS flatbands can become stronger than both the bandwidth and disorder broadening, providing a setting for correlated behavior such as flatband ferromagnetism. We expect PDS flatbands can be experimentally realized in a range of readily available gapped vdW materials such as monolayer transition metal dichalcogenides, e.g. WSe$_2$. 
\end{abstract}

\maketitle

van der Waals (vdW) heterostructures have become a powerful platform to tailor the electronic properties of materials~\cite{Geim2013, Novoselov2016, Song2018}. One case in point is moir\'e superlattices, formed when two vdW materials are stacked and twisted. In such moir\'e materials, electrons and other quasiparticles experience slowly varying (emergent) effective periodic potentials. Even when the potentials are relatively weak (as compared with the kinetic energy of electrons in each layer), a wide variety of phenomena can be realized that include for e.g., emergent electronic bandgaps~\cite{Hunt2013}, moir\'e excitons~\cite{Yu2017, Jin2019, Tran2019, Seyler2019, Alexeev2019}, Hofstadter spectra~\cite{Hunt2013,Ponomarenko2013, Dean2013}, as well as topological bands~\cite{Song2015, Tong2017}, to name a few. However, when top and bottom layers couple strongly, extreme bandstructure reconstruction takes effect, allowing nearly flat electronic bands to form at magic~\cite{MacDonald2011,Santos2007} or low-twist angles~\cite{Wu2018,Wu2019}. These provide a vdW venue to realize correlated behavior, with intense interest sparked by reports of correlated insulating behavior and superconductivity in moir\'e materials under such conditions~\cite{Cao2018a, Cao2018b, Chen2019}. 

However, achieving good twist angle control over moir\'e supelattices can often be experimentally difficult; this can be further complicated by lattice reconstruction that arise at low twist angle~\cite{Alden2013}. Good twist/registration control becomes critical given that extreme bandstructure engineering typically occurs only at specific ``magic'' twist angles or in a small range of low-twist angles~\cite{MacDonald2011,Santos2007,Wu2018,Wu2019}. Recently, an inverted electrostatic strategy --- patterned dielectric superlattices (PDS) --- wherein a dielectric material is patterned into a superlattice and placed on top of a gate electrode (Fig. 1a) has been experimentally demonstrated in graphene~\cite{Forsythe2018}, producing high quality superlattices. In these, dielectric contrast between patterned hole and dielectric substrate material enables the gate electrode to sustain a spatially modulated superlattice potential (Fig. 1b). Perhaps most remarkable is that the PDS devices maintained an ultrahigh mobility (where the patterned dielectric did not significantly degrade the device performance~\cite{Forsythe2018}), as well as high gate electrode tunability that could turn electrically on and off the superlattice potential.

\begin{figure}[t!] 
\includegraphics[width=1\columnwidth]{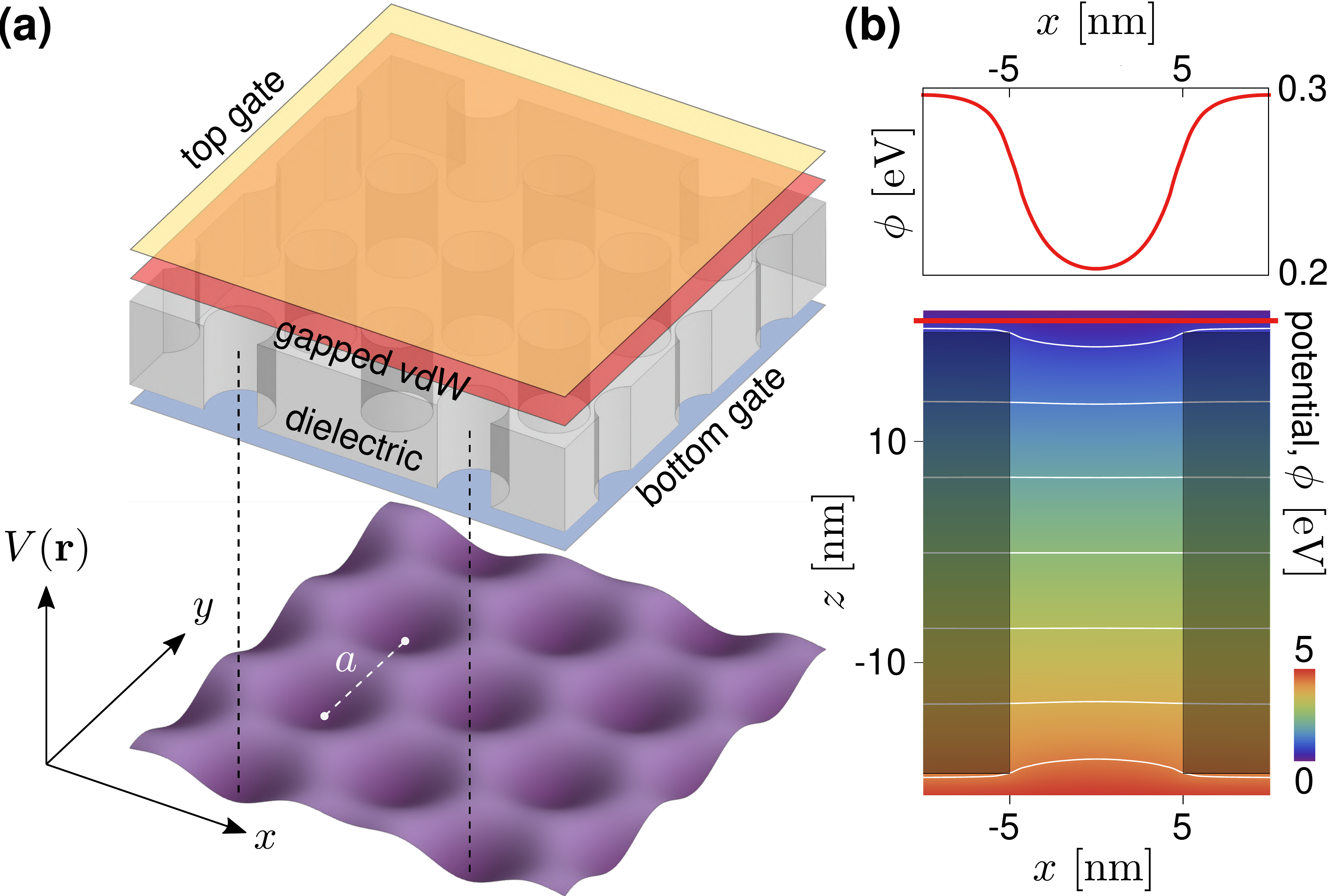}
\caption{
(a) Schematic of a triangular patterned dielectric superlattice (gray) and the spatial potential profile (purple) sustained for the target gapped van der Waals material (red). Here the top gate (yellow) and back gate (blue) sustain a potential drop in the out-of-plane direction. The spacing between two neighboring holes is $a$.
(b) Electrostatic potential for a single hole inside the dielectric substrate (gray) in between top and bottom gates, see {\bf SI} for full details.
The top panel shows the electric potential across the red line (plane of the vdW material) displayed in the bottom panel. Thin white lines indicate equi-potential contours. Here we have used potential difference between the two (top and bottom) gates of $5\, {\rm eV}$ for illustration giving a superlattice potential amplitude ($V_s$ see Eq.~\ref{eq:hamiltonian}) of about 10 meV; we note these correspond to electric fields smaller than the dielectric breakdown voltage of hBN.}
\label{fig1}
\end{figure}

\begin{figure*}[!]
\includegraphics[scale=0.225]{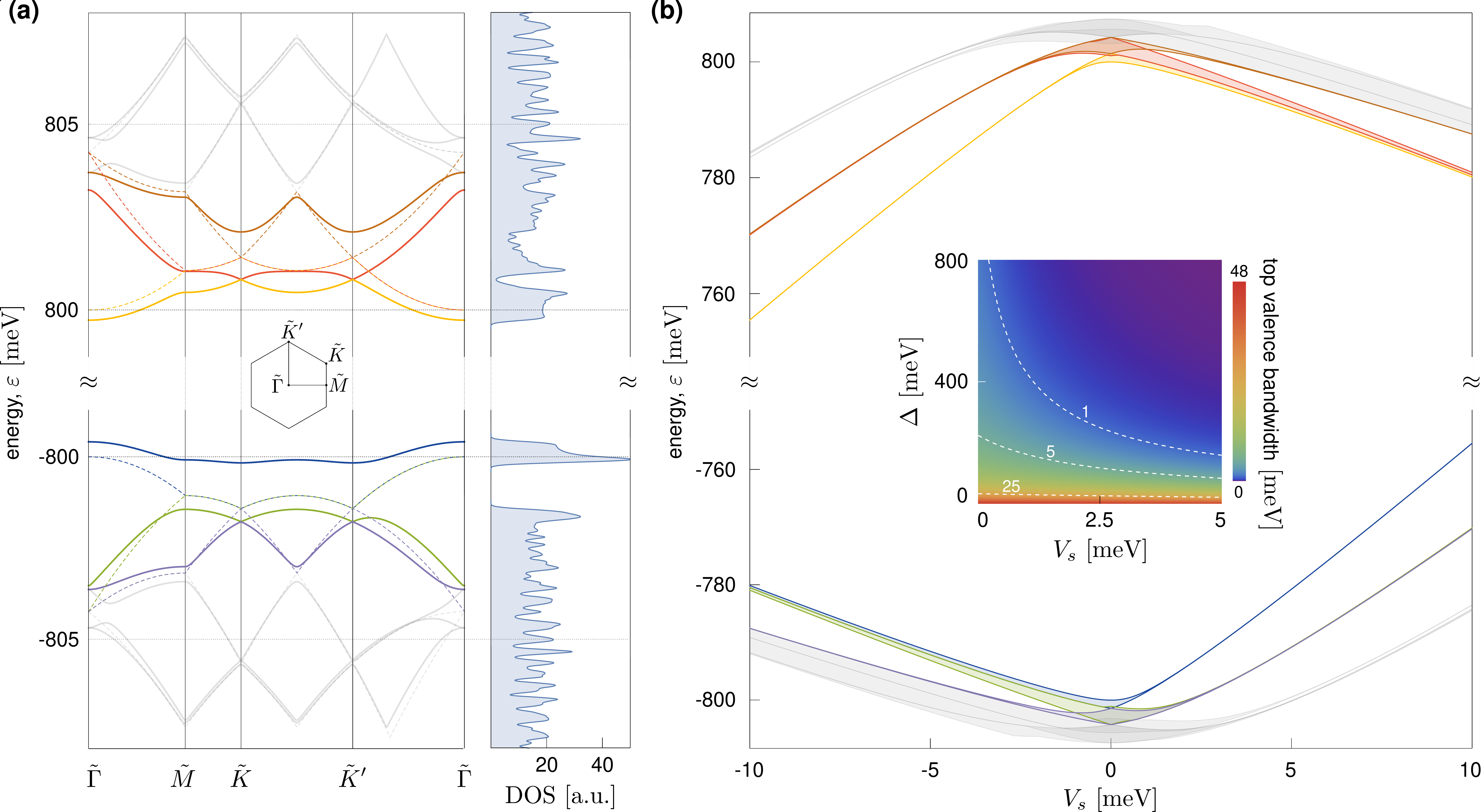}
\caption{
(a) Superlattice miniband dispersion (left) and the corresponding density of states (right) for vdW PDS with monolayer WSe$_2$. Here the electronic structure is described by Eq.~(\ref{eq:hamiltonian}) with $v \hbar = 3.94\,{\rm eV \AA}$, $\Delta = 0.8\,{\rm eV}$; the substrate is a triangular patterned dielectric superlattice with $a = 20\,{\rm nm}$ [see Fig.~(\ref{fig1}a)]. Solid (dashed) lines are for $V_s = 0.5\,{\rm meV}$ ($V_s = 0$) [see Eq.~(\ref{eq:hamiltonian})], and the inset shows the mini Brilluoin zone defined by the superlattice.
(b) Miniband minima and maxima that indicate its bandwidth are plotted as a function of $V_s$ for 12 mini-bands close to the intrinsic band gap. Here solid colored lines indicate miniband maxima and minima, whereas shaded region indicate region in between maxima/minima in the minibands. All other parameters (aside from varying $V_s$) are the same with those in (a). The inset shows the top valence bandwidth as a function of $\Delta$ [with $v \hbar $ and $a$ fixed to the values indicated in (a)].
}
\label{fig2}
\end{figure*}

Here we argue that PDS can be pushed into the strong coupling regime, where the superlattice potential ($V_s$) can exceed the kinetic energy of the electrons in a gapped vdW material. As we describe below, in this regime, flat electronic minibands can be achieved and are highly gate-tunable. Strikingly, these flatbands do not require fine tuning of twist angle. 
Indeed, the bands become flat as gate voltage is switched on, achieving small bandwidths even for modest superlattice potentials (Fig.~\ref{fig2}). This stands in contrast to strategies using moir\'e superlattices that only feature flat bands at magic or low twist angles~\cite{Wu2018,Wu2019,MacDonald2011,Santos2007}. 

PDS flatbands can be achieved in gapped vdW materials such as the transition metal dichalcogenides. The key ingredients are a large bandgap as well as a large effective mass. The former ensures that conduction and valence bands do not cross when $V_s$ is applied, and the latter gives a small initial kinetic energy of the electrons. These enable the superlattice potential to confine the electrons effectively and form flat minibands. 
A particularly good candidate for PDS flatbands is WSe$_2$, wherein high mobility ($\gtrsim 3\times 10^{4}$ cm$^2/$Vs) samples have been isolated~\cite{KimAPS}; these have mean free paths of several hundred nanometers~\cite{Supp}. For WSe$_2$, we anticipate flat electronic minibands that are well separated can be achieved for modest $V_s$ and superlattice periods. Interestingly, PDS yields bunches of flat bands that proliferate throughout the parameter space. For example, for $V_s >0$, we find that in addition to the top most valence miniband which is a single non-degenerate flatband (per valley), the other lower (valence) minibands also bunch up into flatband bundles. 
Similar flatband bundles can be found for $V_s<0$, as well as in the conduction band.

Perhaps the most exciting aspect of PDS flatbands is the possibility of direct gate access to correlated behavior. As an illustration, we consider ferromagnetism in a PDS flatband; this is in close analogy with quantum-hall ferromagnetism. In such a case, we find a ferromagnetic state, 
where the spins in either valley $K/K'$ are favored, can be achieved for readily available samples and superlattice gate voltages. 

{\it Gate tunable flat minibands --} We begin with a discussion of 
the PDS scheme using monolayer transition metal dichalogenides (TMD). 
The PDS that we consider here (and that can be experimentally fabricated) are of fairly long-wavelength $a \gtrsim 20 \, {\rm nm}$. 
The resulting superlattice bandstructure can be captured by an effective $k\cdot p$ model expanded close to the $K$ and $K'$ points as 
\begin{align}
\mathcal{H} & =v\hbar (\sigma_x k_x + \xi \sigma_y k_y) + \Delta \sigma_z+V_s\sigma_0 f(\vec r), 
\label{eq:hamiltonian}
\end{align} 
where $v$ is the velocity, $\sigma_{x,y,z}$ are the Pauli matrices, $2\Delta$ is the bandgap of the TMD, $\xi = \pm 1$ for $K$ and $K'$ valleys, $V_s$ is the amplitude of the superlattice potential applied by the PDS scheme in Fig.~\ref{fig1}, and $f(\vec r)$ is the spatial pattern of the superlattice potential.  
Here we have suppressed spin indices since the bandstructure physics we discuss does not differentiate between spin/valley species (i.e. the superlattice potential is a scalar potential); spin can be included in a straightforward fashion that does not affect the physics we discuss below. Further, we note that the particularly large Ising spin splitting ($\sim 100$s of meV) in the valence bands of many TMD materials, effectively lock the valley and spin in the valence bands making the model in Eq.~(\ref{eq:hamiltonian}) a good descriptor of the low-energy mini-bandstructure.

As shown in Fig.~\ref{fig1}b, dielectric inhomogeneity in the substrate material can enable electric control of a spatially inhomgeneous potential with variations on the order of the hole width (e.g., several tens of nanometers) \cite{Forsythe2018}. Here gray indicates a dielectric (shown SiO$_2$ with $\epsilon =4$) and empty indicates air ($\epsilon =1$). As shown, when a gate potential is applied, the dielectric contrast yields a spatially modulated potential even in the vdW material layer (red) with potential amplitudes $V_s$ that can be switched on and off electrically \cite{Forsythe2018}. Fig.~\ref{fig1}b was plotted using a numerical solution for a generalized Poisson equation in a spatially inhomogeneous dielectric environment (see Supplementary Information, {\bf SI}). 

To illustrate the (flat) band structure engineering in the PDS of vdW materials, we first use a triangular superlattice (see other lattices in {\bf SI})
so that 
\be
f(\vec r) = \sum_{i=1,2,3}  2 {\rm cos} (\vec G_i \cdot \vec r + \phi_i),  
\label{eq:superlattice}
\ee
where $\vec G_i$ are the triangular superlattice wave vector oriented $60^\circ$ relative to each other, $|\vec G_i| = 2\pi/a$, and $\phi_i$ is a relative phase. For simplicity, we will set $\phi_i=0$ in the following. It does not qualitatively affect the physics we discuss below. As expected, the superlattice folds the original TMD bands into a series of superlattice minibands (band folding is illustrated for $V_s=0$ as the light dashed lines in Fig.~\ref{fig2}a). Here the bandstructure is plotted in a superlattice defined mini Brilluoin zone (MBZ), where the tilde symbol indicates the MBZ. When $V_s$ is switched on, Bragg scattering mixes the bands and produces a mini-band structure (colored lines and gray lines with $V_s = 0.5\,{\rm meV}$) in both the conduction band (positive energies) and the valence band (negative energies) as shown in Fig.~\ref{fig2}a. We have colored only the first three conduction and valence minibands so as to focus our discussion on them; in solving for the mini-bandstructure a set of 
162 mini-bands are employed 
to ensure convergence of the lower bands (we only show 12 minibands) that we focus on in the main text. 
Physically, the inclusion of many minibands in our numerical calculation is to 
to capture the physics of the extreme bandstructure reconstruction regime in which the superlattice strength $V_s$ is larger than that of a kinetic energy of the electrons. 
Here we have used material parameters corresponding to WSe$_2$ with $v \hbar = 3.94\,{\rm eV \AA}$, $\Delta = 0.8\,{\rm eV}$ \cite{Xiao2012}.

Strikingly, a clear gap between the top most (valence) miniband (blue, Fig.~\ref{fig2}a) and the rest of the minibands opens up. This is distinct from that expected from graphene PDS \cite{Forsythe2018} where a $\pi$ berry phase prevents backscattering and gap opening; in the case of graphene PDS secondary Dirac cones form at the MBZ corners~\cite{Forsythe2018, Yankowitz,Hunt2013,Ponomarenko2013, Dean2013}. In contrast, the wavefunction (AB-sublattice pseudospinor) in TMD close to the band edge has weight mostly on a single sublattice (aligned to a pole in the Bloch sphere), allowing maximal backscattering and mini-gap formation. 

Due to the large $\Delta$ in TMDs, minigap opening squeezes the top most valence miniband creating a confined energy window for it to exist with a sharp density of states (shaded blue in right panel of Fig.~\ref{fig2}a). Indeed, as $V_s$ is increased, the top most (valence) mini-band is further flattened (Fig.~\ref{fig2}b) with very narrow band-widths ($\ll 1\,{\rm meV}$) achievable with modest applied gate voltage creating a nearly {\it flat band}. To see this, we have plotted the maximum and minimum energy in each miniband (solid lines Fig.~\ref{fig2}b, this indicates the bandwidth) with the colors used corresponding to the colored 3 conduction and colored 3 valence minibands of Fig.~\ref{fig2}a; the shaded region between miniband maxima and minima are shaded in corresponding colors. We note that the flat top most (valence) miniband is well separated from other bands with the large $\Delta \sim {\rm eV}$ to the minibands in the conduction band, and gate tunable minigaps to the other minibands in the valence band (that can reach tens of meV); the minigaps increase with applied gate potential~\cite{Supp}. 

One unusual feature is that the application of superlattice PDS also renormalizes the effective bandgap between the conduction and valence band states, with changes in effective bandgap of up to $60 \, {\rm meV}$ for the largest superlattice potential amplitude shown in the figure (see Fig.~\ref{fig2}b). This arises from the significant miniband reconstruction. Indeed locally in real space, the electrons feel large variations in superalattice potential that range from $+6V_s$ at the peak to $-6V_s$ in the troughs. This large peak-to-trough difference enables electrons to be confined locally to produce a flatband structure (Fig.~\ref{fig2}a).  

The blue band is not the only flat band that occurs in PDS. For example, at sufficiently high superlattice potential ($V_s>0$), both green and purple minibands in the valence band flatten out. Unlike the blue band they tend to stick together. In fact, the PDS scheme yields sets of flat minibands in the valence band (some of which are nearly degenerate with each other forming bunched bundles of bands) with severely compressed bandwidths. Each of these bundles of flat bands are well separated from each other with large electrically tunable minigaps (in Fig.~\ref{fig2}b, we show three flat miniband bundles in the valence band at $V_s >0$). 

We note that the PDS scheme, when applied to gapped Dirac materials such as WSe$_2$, naturally breaks particle-hole symmetry with (positive and negative energy) minibands exhibiting contrasting behavior (see Fig.~\ref{fig2}a). For example, when $V_s>0$ the top (valence) miniband (blue) gets squeezed into a single non-degenerate flat miniband and is well separated from the other minibands, while the bottom (conduction) miniband (yellow) adheres closely to the next miniband (red); this displays a particle-hole asymmetric behavior. Symmetry in the miniband structure, however, is restored when {\it both} $V_s \to - V_s$ and $\varepsilon_{\vec k} \to - \varepsilon_{\vec k}$ are interchanged (see detailed discussion in {\bf SI}). 

This unconventional feature allows the type of flat mini-bandstructure to be tuned by gate voltage. When $V_s>0$  the top (valence) miniband (blue) is a single non-degenerate flat band (per valley). However, when $V_s<0$ this same blue miniband, while flattening out, adheres closely to the green band (Fig.~\ref{fig2}b); when large enough gate voltage is applied, they form a close pair of flatbands that bunch together. This flatband bundle has a width $\sim 1\,{\rm meV}$. Even though both green and blue minibands stay together, nevertheless, we find that they are  
separated by a very small energy gap. Indeed, close to $\tilde{K}, \tilde{K}'$ points, the blue and green miniband structure (for $V_s<0$) bands resemble Dirac cones (with an extremely small gap; we estimate the gaps to be of order several $\mu\,{\rm eV}$, see {\bf SI}). We note that away from the small minigaps, the minibands spectra mimic the flattened Dirac bands in twisted bilayer graphene~\cite{MacDonald2011}; interestingly, the chirality of electrons at $\tilde{K}, \tilde{K}'$ points are the same~\cite{Supp} mirroring the behavior found in twisted bilayer graphene.

As a result, PDS enables to achieve multiple types of flatbands via in situ gate voltage tuning (e.g., from single non-degenerate flat-miniband for positive $V_s$ to Dirac-cone-like for negative $V_s$). This unusual asymmetric behavior (for both conduction/valence bands and opposite signs of $V_s$) arises from the gapped Dirac pseudo-spinor form of their wavefunctions (such asymmetry does not arise for a simple massive two-dimensional electron gas). Indeed, the role of the TMD pseudospinor texture is further evidenced by how PDS induced Bragg scattering also changes the winding of the gapped Dirac pseudo-spinor wavefunctions. This dramatically reconstructs the Berry curvature distribution in each of the minibands (for discussion of miniband Berry curvature, see {\bf SI}). 

One of the most attractive feature of PDS flat bands, is that they do not occur at fine-tuned gate voltages or superlattice wavelength.
The larger the superlattice wavelength, the smaller the applied superlattice potential needed to flatten the PDS minibands.
This strategy stands in stark contrast to the flat bands found in moir\'e superlattices where specific ``magic'' twist angles between layers are required (in the case of twisted bilayer graphene) or low twist angles required in other moir\`e superlattice strategies. The key to achieving flat bands in PDS is the large $\Delta$ as well as low velocities. Indeed, as shown in the inset of Fig.~\ref{fig2}b, 
when we fix $v \hbar$ and $a$ (same parameters as panel a),
the larger the $\Delta$, the smaller the bandwidth at a given superlattice potential $V_s$. For WSe$_2$, we find $\Delta = 800\,{\rm meV}$ allowing very small bandwidths to be achieved even for small $V_s$ applied, e.g., the bandwidth is $\sim {\rm \mu eV}$ when $V_s = 5 \,{\rm meV} $. We note, parenthetically, that when $\Delta$ is small so that $V_s$ applied is on order $\Delta$, the minibands in the conduction and valence band can start to mix, vastly complicating the miniband structure and making the conditions for flat bands in PDS hard to achieve. 

We anticipate that the PDS scheme can be applied to other van der Waals materials with large bandgaps. For example, we have computed the miniband structure for a range of TMD materials and have found that well separated sets of flat minibands generically occur -- see {\bf SI} for full band structure. Further, other superlattices can also be easily implemented and produce qualitatively the same results: we have computed the minibands for square as well as hexagonal superlattices as well and find similar well separated sets of flat minibands~\cite{Supp}. This versatility with superlattice structure provides the ability to study flatbands and its concomittant interaction effects in other types of lattices which can have a different type of symmetry (that may be different from the underlying lattice).  

\begin{figure}[t!]
\includegraphics[width=\columnwidth]{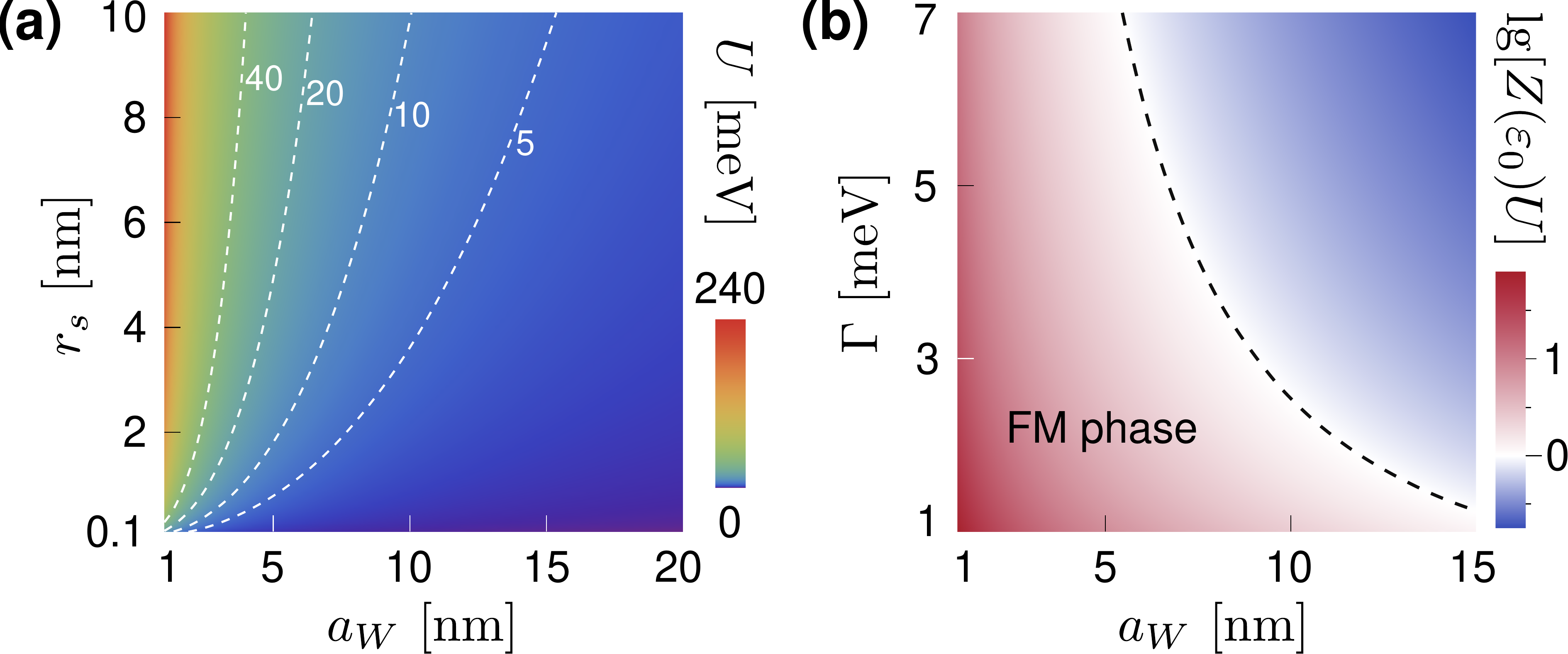}
\caption{
(a) Interaction energy $U$ as a function of the extent of electronic wavefunction in the flat band, $a_W $, and effective screening radius, $r_s$. Dashed contour lines indicate lines of constant $U$ and are in units of meV. 
(b) Dimensionless $Z(\mu_0=\varepsilon_0) U$, see Eq.~(\ref{eq:stoner}) indicating propensity for the ferromagnetic instability with $Z(\mu_0=\varepsilon_0) = [ \sqrt{2\pi \Gamma^2} ]^{-1}$ and taken at a fixed $r_s = 5 \, {\rm nm}$ for illustration. For $Z(\mu_0) U >1$ (boundary denoted by dashed line), the system enters a ferromagnetic instability. Here lg denotes ${\rm log}_{10}$.
}
\label{fig3}
\end{figure}

{\it Flatband ferromagnetism --- } Perhaps the most striking consequence of PDS flat bands is the ability to enhance correlation effects. This is because the extremely small bandwidth of single-particle mini-bandstructure of PDS ($\ll 1\, {\rm meV}$) quenches the kinetic energy of the electrons. As a result, other energy scales such as that arising from electron-electron interactions
can determine the behavior of the electronic system. For clarity, in the following, we will focus on the top-most valence miniband for $V_s>0$ which is well separated from the other superlattice minibands. Further, we note that due to the large Ising splitting in TMDs, the spin and valley degree of freedom are locked; at the non-interacting level, this band has only a two-fold degeneracy with spin up and spin down occurring (original) valley at $K$ and $K'$. 

To proceed, we first estimate the strength of electron-electron interactions in the superlattice by modeling the average interaction energy of electrons confined in the troughs of the superlattice potential as \cite{MacDonald1986,Nomura2006}
\be
U = \sum_{\vec q} \frac{2\pi e^2}{\epsilon (q + q_s)} {\rm exp} (-q^2 a_W^2), 
\label{eq:interactions}
\ee
where $q_s = r_s^{-1}$ is the effective inverse screening length, and $a_W$ is the extent of the PDS flatband electronic wavefunction 
confined in the troughs of the superlattice. Here we have chosen a simple Gaussian ansatz for the extent of the wavefunction; other models do not qualitatively affect the results we discuss below. In Fig.~\ref{fig3}a, we plot the strength of $U$ as a function of both the screening length $r_s$ as well as $a_W$ taking $\epsilon =4$. 
This yields large interaction energies of order several tens of meV for a wide range of $r_s$ and $a_W$. For WSe$_2$, we estimate effective $r_s $ of a few nanometers arising either from intrinsic screening of the electron gas or from proximal gates, see discussion in {\bf SI}. To estimate the extent of the electronic wavefunction $a_W$, we employ a variational approach on the Gaussian wavefunction ansatz. This yields typical $a_W$ that decreases for increasing superlattice potential, reaching a fairly confined state of $\sim 5-10\, {\rm nm}$ even for modest applied superlattice potential, see {\bf SI}. 

Fig. 3a indicates that $U$ can dominate over the extremely small non-interacting PDS flatband bandwidths (see e.g., Fig.~\ref{fig2}). As a result, we anticipate interaction effects can become significant. For example, large exchange energy can drive the spin-degenerate flatband system into a ferromagnetic state. To describe this, we use a simple mean-field model for the energy density $\mathcal{E}_{\rm MF}$ of a PDS flat miniband with spins indexed $s = \uparrow, \downarrow$ as 
\be
\mathcal{E}_{\rm MF} =  \sum_s \Big[\int^{\mu_s} \text{d} \varepsilon \, \varepsilon Z(\varepsilon) - \frac{U}{2} n_s^2\Big], 
\label{eq:meanfieldenergy}
\ee
where $\mu_s$ is the chemical potential of spin $s$, and $\varepsilon$ is the band energy, and $U$ is the strength of the exchange energy and can be estimated by Eq.~\ref{eq:interactions} [for full description, see {\bf SI}]. Here we have used a simple Gaussian 
$Z(\varepsilon) = A \exp [ - (\varepsilon - \varepsilon_0)^2 / 2\Gamma^2 ]$ to model a broadened spectral weight of the flatband, and the density of each spin species is $n_s = \int^{\mu_s} {\rm d} \varepsilon Z(\varepsilon)$. The broadening, $\Gamma$, can be induced by a number of different processes for e.g., via disorder. Other forms of the spectral function do not affect the qualitative conclusions we discuss below. Here $A = [ \sqrt{2\pi \Gamma^2} ]^{-1}$ is a normalization constant. 

For small $U$, the two spin species are degenerate with $\mu_\uparrow = \mu_\downarrow = \mu_0$ is the spin chemical potential. However, for large enough $U$ this spin symmetry can become broken. To see this, we first write $n_s = n_0 \pm \delta n$ where $2 n_0 = n_\uparrow + n_\downarrow$. Expanding Eq.~(\ref{eq:meanfieldenergy}) to $ \mathcal{O} (\delta n^2)$ we obtain the energy density for the flatband as 
\be
\mathcal{E}_{\rm MF} = 
\mathcal{F}_0
+ \delta n^2 [ 1/ Z (\mu_0)- U]  + \mathcal{O} (\delta n^4),
\label{eq:stoner}
\ee
where $\mathcal{F}_0 = 2 \int^{\mu_0} \text{d} \varepsilon\,  \varepsilon Z(\varepsilon)- U n_0^2$ is the energy in the symmetry unbroken phase, and the $\delta n^2$ term describes the energy cost to imbalance the spin species. Crucially, an instability in the spin up/down population is induced when the coefficient of the second term is negative $Z(\mu_0) U > 1$. 

The conditions for symmetry breaking depend on a competition between the broadening 
and the exchange energy. In our model above, this is parameterized by $r_s$, $a_W$, as well as the broadening energy/width $\Gamma$. To illustrate this, we plot $Z ( \mu_0 = \varepsilon_0 ) U$ in Fig.~\ref{fig3}b; this is the maximal value that is obtained at half-filling. From Fig.~\ref{fig3}b, we can see the competition clearly: for larger (smaller) $\Gamma$, we require more (less) localized electron states, or larger (smaller) exchange energy to enter the broken symmetry phase. 
%Crucially, even 
Taking $r_s = 5 \, {\rm nm}$ as a demonstration, we find large $Z ( \mu_0 = \varepsilon_0 ) U$ for a wide swathes of the $\Gamma$-$a_W$ parameter space (Fig.~\ref{fig3}b). We estimate that $a_W \sim 5-10\, {\rm nm}$ can be achieved by modest superlattice potentials in the PDS scheme. At low temperatures, broadening is typically dominated by disorder~\cite{Adam2007,Skinner2013,Skinner2014}. Taking values for high quality WSe$_2$~\cite{KimAPS}, we estimate disorder-induced broadening $\Gamma_{\rm dis} \sim$ of order several ${\rm meV}$ are available in high quality present day samples, see {\bf SI}. As a result, we expect that the conditions for realizing a ferromagnetic state using PDS flatbands can be attained in WSe$_2$. Interestingly, this analysis is not confined to half-filling $\mu_0=\epsilon_0$. The ferromagnetic instability can occur at a variety of chemical potentials so long as the criterion is satisfied $Z (\mu_0) U >1$, see below. 

{\it Discussion ---} There are numerous probes of the extreme mini-bandstructure reconstruction we discuss here. For example, we anticipate a lightly hole doped WSe$_2$ PDS to exhibit a dramatic change in its low-frequency (THz) optical absorption characterisitcs as $V_s>0$ is switched on. While at $V_s=0$ such a sample will exhibit a Drude peak around $\omega =0$, when $V_s>0$ is switched on, the Drude peak will diminish as the topmost (valence) miniband flattens, and exhibit additional sharp THz absorption peaks corresponding to transitions between the sets of flat minibands (in the valence band).

Similarly, we expect dual-gate control to enable control of both superlattice potential as well as the filling of the minibands. Such filling control can enable to probe both the metallic as well as the insulating states induced by the superlattice potential. When pushed into the regime where flatbands exhibit ferromagnetism, the spins (locked to the valley) split; at half-filling of a flatband this would exhibit a ferromagnetic insulating state. Interestingly, ferromagnetism can also occur away from half-filling so long as $Z (\mu_0) U >1$. As a result, the ferromagnetic system can be metallic. Crucially, we note that the minibands exhibit non-zero Berry curvature distribution (see details in {\bf SI}). As a result, when the spins/valleys are split (and $\mu_0$ is away from half-filling) we anticipate that a (charge) anomalous Hall effect can ensue. 

In summary, PDS in gapped vdW materials provide a venue to realize flatbands without the need for sensitive twist angle alignment or stacking arrangement. Instead, extreme bandstructure engineering arise directly from gate-controlled superlattice potentials yielding flatbands when gate voltage is applied; when gate bias vanishes, the PDS system remains as an ordinary TMD system. As a result, PDS flatbands afford considerable electrical control over the character of their electronic excitations (e.g., miniband-width, miniband-gap, interaction energy), as well as an on/off switch for flatbands. While we have focussed on ferromagnetic ordering, highly localized electrons can also exhibit other types of magnetic ordering including antiferromagnetic order as well spin liquids (for a recent discussion see e.g. Ref~\cite{Wu2018}); having two extremely flat bands separated by a small gap may provide a possible venue for unusual excitonic correlations~\cite{Rice1967}. Given the large variety of available superlattice structures that can be patterned using PDS, the characteristic length scales for PDS (several tens of nanometers), as well as the exposed nature of the surface states, such heterostructures may provide a one-stop platform to realize and spatially probe exotic ordering.  

{\it Acknowledgements --- } We are grateful for useful conversations with Valla Fatemi, Ata\c c Imamo\u glu, Brian Skinner, Javier Sanchez-Yamigishi, and Noah Yuan. This work was supported by the Singapore National Research Foundation (NRF) under NRF fellowship award NRF-NRFF2016-05, a Nanyang Technological University start-up grant (NTU-SUG), and Singapore MOE Academic Research Fund Tier 3 Grant MOE2018-T3-1-002.

\clearpage

\newpage

\setcounter{equation}{0}
\renewcommand{\theequation}{S-\arabic{equation}}
\renewcommand{\thefigure}{S-\arabic{figure}}
\renewcommand{\thetable}{S-\Roman{table}}
\makeatletter
\renewcommand\@biblabel[1]{S#1.}
\makeatother

\setcounter{figure}{0}
\twocolumngrid

%\renewcommand\thefigure{\thesection.\arabic{figure}}    
%\section{A nice appendix}
%\setcounter{figure}{0}    

\section*{Supplementary Information for ``Gate-tunable flat bands in van der Waals patterned dielectric superlattices''}
\appendix

\section{Superlattice band structure}

For a 2D trianglular superlattice, we model the Hamiltonian as
$\mathcal{H} = \mathcal{H}_0 + V (\vec r)$, with
\begin{equation}
\mathcal{H}_0 = v \hbar (\sigma_x k_x + \xi \sigma_y k_y) + \Delta \sigma_z , 
\end{equation} 
and 
\begin{align}
V (\vec r) = V_s\sigma_0 \sum_{j=1}^3  2 {\rm cos} (\vec G_j \cdot \vec r) ,
\,
{\vec G}_j = \frac{4\pi}{3a} \Big( \cos \frac{j \pi}{3}, \sin \frac{j \pi}{3} \Big) .
\label{eq:potential}
\end{align} 
Here $ {\vec G}_j$ are the triangular superlattice reciprocal vectors oriented $60^\circ$ relative to each other.
The band structure is obtained by diagonalizing the Bloch Hamiltonian, whose $(pq, mn)$-th $2 \times 2$ block reads
\begin{align}
\big[ H (\vec k) \big]_{pq, mn} = &~ \la \vec k_{pq} | \mathcal{H}_0 | \vec k_{mn} \ra 
\nn
& + 
V_s  \sum_{j=1,2,3} \la \vec k_{pq} | 2 \cos (\vec G_i \cdot \vec r) \sigma_0 | \vec k_{mn} \ra ,
\end{align}
where $\vec k_{pq} = \vec k + p \vec G_0 + q \vec G_1$ with $\vec k $ within the first Brillouin zone, $p,q\in \mathbb{Z} $, and $| \vec k_{pq}  \ra$ is the plane wave basis. The same applies to $\vec k_{mn}$ and $| \vec k_{mn}  \ra$
 For any $(pq, mn)$ block, it is coupled to the other 6 blocks: $(pq\pm1, mn)$, $(pq, mn\pm1)$, $(pq-1, mn+1)$ and $(pq+1, mn-1)$. 

Numerically, we have to truncate the infinitely large matrix $ H (\vec k)$ into a finite sized $H_N (\vec k)$ in which $p,q \in [-N, N]$, and $H_N (\vec k)$ has a finite dimension $2 (2N+1)^2$. Throughout the paper, we find $N=4$ (162 mini-bands in total) is sufficient to ensure the dispersion for the minibands close to the intrinsic gap converge.

\section{Electric potential from generalized Poisson equation\label{appx:GPE}}

For a non-uniform dielectric environment, we have the generalized Poisson equation
\begin{equation}
\nabla \cdot \vec D (\vec r) = 
- \nabla \cdot [ \epsilon (\vec r) \nabla \phi (\vec r) ]  = 4\pi \rho_\text{free} (\vec r) .
\label{eq:gpe}
\end{equation}
With the knowledge of $\epsilon (\vec r)$, and the values of $\phi (\vec r)$ at the top and bottom gates, we can solve $\phi (\vec r)$ using Eq.~(\ref{eq:gpe}).

\begin{figure}[t!] 
\includegraphics[width=0.72\columnwidth]{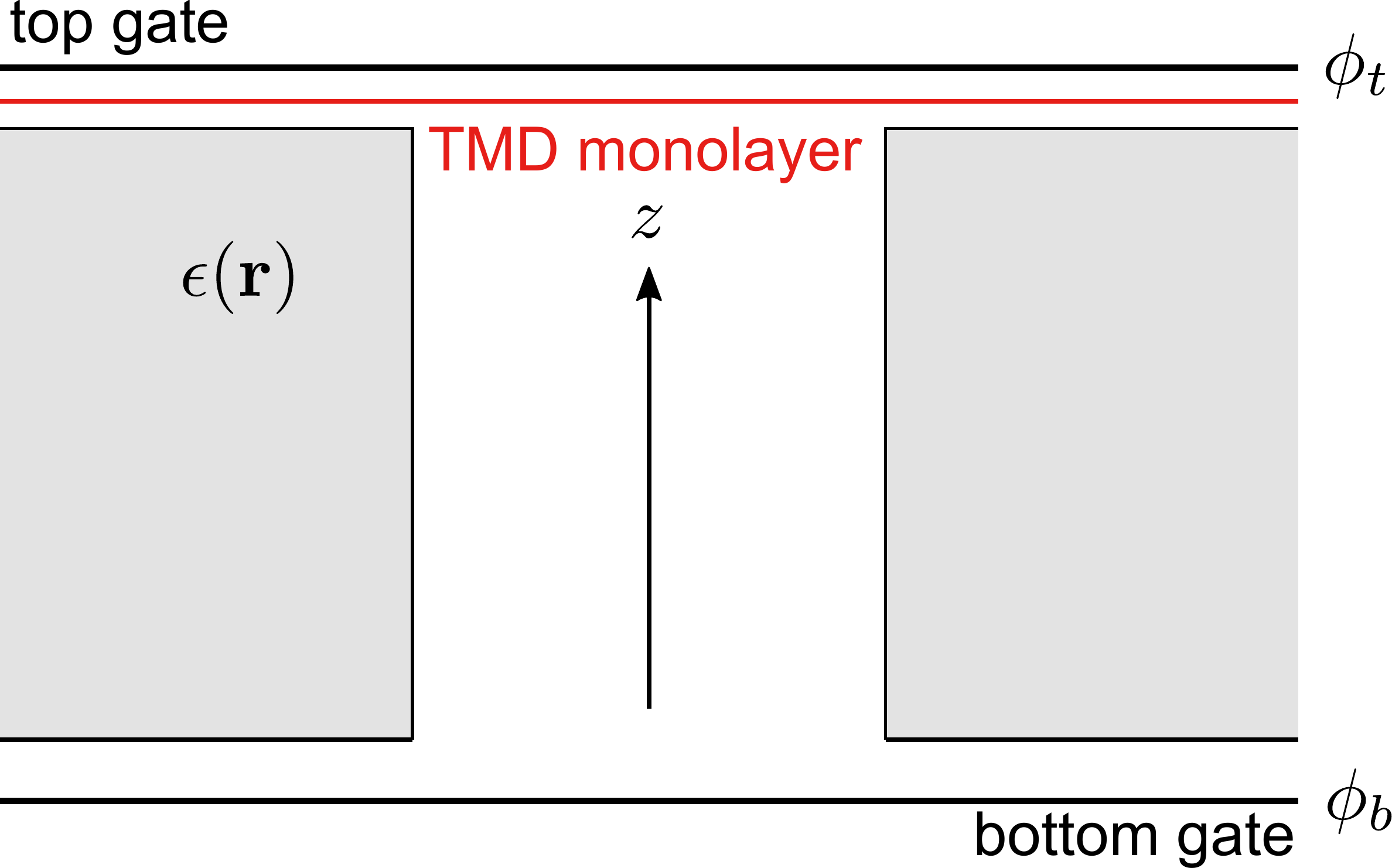}
\caption{The inhomogeneous dielectric environment simulated in Fig.~\ref{fig1}b. Gray regions are SiO$_2$ substrate and the white region denotes an air hole. A TMD monolayer lies in between the substrate and the top gate.
Parameters used in the simulation: the distance (exaggerated here for illustration) between the top/bottom gate and the top/bottom of the SiO$_2$ substrate is $3\,{\rm nm}$; the thickness of the SiO$_2$ substrate in the $z$-direction is $40\,{\rm nm}$; the radius of the air hole within the SiO$_2$ substrate is $5 \,{\rm nm}$; $ \phi_b - \phi_t = 5 \, {\rm eV} $.
}
\label{fig-s1}
\end{figure}

We now investigate the electric potential in the configuration illustrated in the Fig.~\ref{fig-s1}. In this configuration, there is a single hole in the SiO$_2$ dielectric substrate, which is in between a top gate and bottom gate. Since we have rotational symmetry in the $x$-$y$ plane, we choose cylindrical coordinates
$(\rho, \varphi, z)$:
\begin{equation}
x = \rho \cos \varphi,
\quad
y = \rho \sin \varphi,
\quad
z = z ,
\end{equation}
expressing $\epsilon (\vec r) = \epsilon (\rho, z)$, $\phi (\vec r) = \phi (\rho, z)$.
Using rotational symmetry, we eliminate the azimuthal angle to obtain the
2D equation:
\begin{equation}
\frac{1}{\rho} \frac{\p}{\p \rho} \lb \rho ~ \epsilon (\rho, z)  \frac{\p \phi (\rho, z)}{\p \rho} \rb + \frac{\p}{\p z} \lb \epsilon (\rho, z)  \frac{\p \phi (\rho, z)}{\p z} \rb = 0  .
\label{eq:gpe2D}
\end{equation}
To proceed we specify $\epsilon (\rho, z)$ in the whole region ($\epsilon = 4$ for gray regions and $\epsilon = 1$ for white regions), and the boundary conditions at the bottom and top gates for $\phi (\rho, z)$ (see Fig.~\ref{fig-s1}, electric potentials for top and bottom gates are set to $\phi_t$ and $\phi_b$, respectively). We can numerically solve the $\phi (\rho, z) $ in the whole region. In so doing, we used {\it Mathematica} NDSolve to solve the differential Eq.~\ref{eq:gpe2D}.

In our simulation, we obtain $V_s \approx 8.3 \, {\rm meV}$ when we set $\phi_b-\phi_t = 5\, {\rm eV}$(see Fig.~\ref{fig1}b). This correspond to an electric field strength $E \approx 0.11 \, {\rm V\,nm^{-1}}$ which is smaller than the breakdown electric field strength ($\sim 0.5 \, {\rm V\,nm^{-1} }$) for hBN.

\subsection{Particle-hole asymmetry}

The bare TMD Hamiltonian $ {\cal H}_0 = v \hbar (\sigma_x k_x + \xi \sigma_y k_y) + \Delta \sigma_z $ has eigen energy $\varepsilon_{\vec k}^\pm = \pm \sqrt{ v^2 \hbar^2  | \vec k |^2 + \Delta^2}$ and eigen states
\begin{equation}
u_{\vec k}^+ =
\begin{bmatrix}
\cos ( \theta_{\vec k} / 2 ) e^{ - i \phi_{\vec k} / 2 }  \\
\sin  ( \theta_{\vec k} / 2 ) e^{ i \phi_{\vec k} / 2  }
\end{bmatrix}
,
\quad
u_{\vec k}^- =
\begin{bmatrix}
\sin ( \theta_{\vec k} / 2 ) e^{ - i \phi_{\vec k} / 2 } \\
- \cos  ( \theta_{\vec k} / 2 ) e^{ i \phi_{\vec k} / 2 }
\end{bmatrix} ,
\end{equation}
where
$ \cos \theta_{\vec k} = \Delta / \sqrt{ v^2 \hbar^2 | \vec k |^2 + \Delta^2}$ and $ \tan \phi_{\vec k} = k_y / k_x $ .
The particle-hole operation ${\cal P} = i \sigma_y {\cal K}$ (${\cal K}$ is the complex conjugation) transforms ${\cal P} u_{\vec k}^+ = u_{\vec k}^-$, i.e., each eigenstate $u_{\vec k}^+$ at energy $\varepsilon_{\vec k}^+$ has a copy ${\cal P} u_{\vec k}^+$ at energy $\varepsilon_{\vec k}^-$. Meanwhile it also satisfies ${\cal P} {\cal H}_0 {\cal P}^{-1} = - {\cal H}_0$. Therefore ${\cal H}_0$ has a particle-hole symmetry.

However, for the superlattice potential $V (\vec r)$ we have ${\cal P} V(\vec r) {\cal P}^{-1} = V (\vec r)$. This breaks the particle-hole symmetry for the overall Hamiltonian, i.e., ${\cal P} {\cal H} {\cal P}^{-1} \neq - {\cal H}$.

Instead, the hamiltonian $\mathcal{H}$ possesses a different type of symmetry. To see this, we define a composite particle-hole and voltage flip operation $\tilde{\cal P} = {\cal P} {\cal M}_z $. Here $\cal P$ is the particle-hole operation (as defined above), and the mirror operation ${\cal M}_z$ flips $V (\vec r) \to - V (\vec r)$ by interchanging the top and bottom gate potential keeping ${\cal H}_0$ intact. Using this composite operation we recover $ \tilde{\cal P} {\cal H} \tilde{\cal P}^{-1} = - {\cal H} $. More explicitly, 
when $V_s \to - V_s$ then $\varepsilon_{\vec k} \to - \varepsilon_{\vec k}$ are interchanged.

\subsection{Mini-band Berry curvature distribution, valley Chern number, and miniband gapped-Dirac-cone}

\begin{figure}[t!]
\includegraphics[width=0.925\columnwidth]{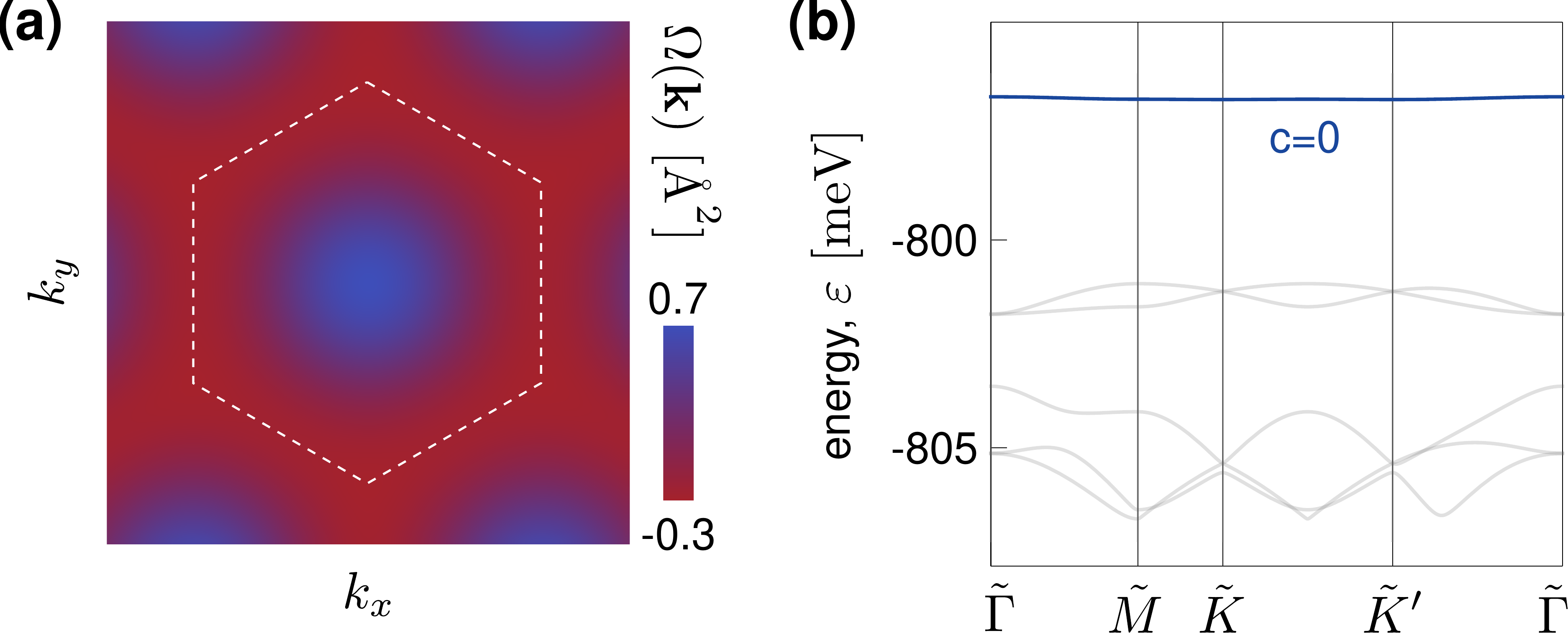}
\caption{
(a) Berry curvature distribution for the top most valence miniband. Dashed hexagons denote the mini Brillouin zone. 
(b) Dispersion for the mini-bands close to the intrinsic band gap, and the valley Chern number for the top most valence miniband. Parameters are the same with those in Fig.~\ref{fig2}a in the main text, except that $V_s = 1.5\, {\rm meV}$.
}
\label{fig-s2}
\end{figure}

\begin{figure}[t!]
\includegraphics[width=\columnwidth]{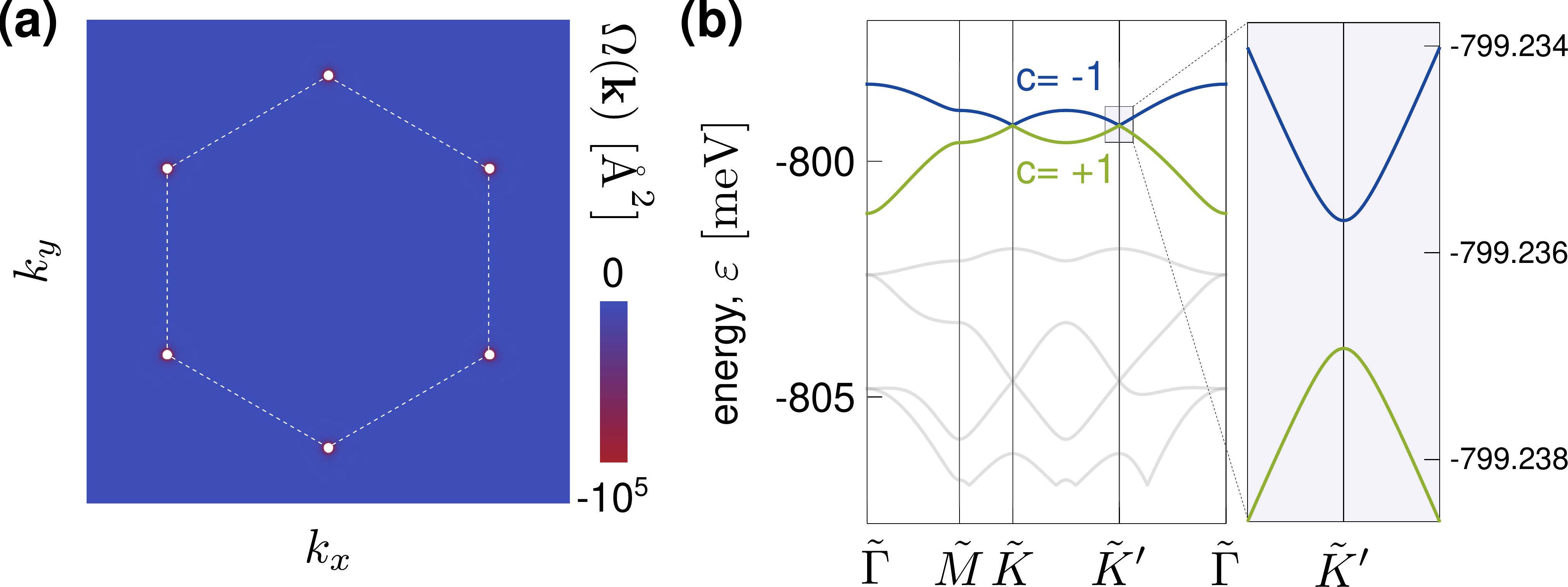}
\caption{Similar to Fig.~\ref{fig-s2}, but with $V_s = -1.5\,{\rm meV}$ with Berry curvature distribution shown for the blue band. Here Berry curvature distribution is sharply peaked close to the corners of the MBZ at $\tilde{K}, \tilde{K}'$ points [clipped white spots denote $|\Omega (\vec k) | > 10^5 \, {\rm \AA^2}$]. Berry curvature values close to the $\tilde{K}, \tilde{K}'$ points are the same sign indicating the pseudospin winding of the same chirality. Berry curvature about $\tilde{K}, \tilde{K}'$ points both have a negative sign. The panel on the right hand side in (b) shows the tiny gap between two nearly degenerate minibands.}
\label{fig-s3}
\end{figure}

The PDS superlattice potential can also modify the winding of the pseudo-spins as well as the miniband quantum geometry. To display this, we plot the Berry curvature distribution in the top most valence mini-band for $V_s>0$ (see Fig.~\ref{fig-s2}a) and
$V_s < 0$ (see Fig.~\ref{fig-s3}a). In what follows, we concentrate on the Berry curvature for minibands from the (original) $K'$ valley with $\xi = - 1$ (see Eq.~\ref{eq:hamiltonian}).
We first note that the Berry curvature around a single bare gapped Dirac cone in the valence band of the TMD (when $V_s=0$)
has a positive sign for $K'$ valley, and monotonically decreases in magnitude away from the Dirac point. 

We now focus on the topmost flat valence miniband when $V_s>0$. Now the (positive) sign of the Berry curvature from the bare gapped Dirac cone in the TMD is reflected in the center of the MBZ (see Fig.~\ref{fig-s2}a). However, PDS induced Bragg scattering close to the MBZ boundaries produces a different winding of the pseudospin and a different sign of Berry curvature close to MBZ boundaries. This reconstructs the Berry curvature distribution. As a result, the net Berry flux through the MBZ for the top most valence miniband is $\mathcal{C} = 0$. 

Next, we turn to the topmost flat miniband when $V_s < 0$. Similar to the case discussed above, the positive sign of the Berry curvature from the bare gapped Dirac cone in the TMD is again reflected in the center of the MBZ. 
However, Bragg scattering at MBZ corners in the top valence miniband for $V_s<0$ is significant and dramatically changes the winding of the pseudo-spinor wavefunction; indeed the Berry curvature distribution is concentrated at the MBZ corners displaying large and negative value (see small white dots). Note that the white dots appear at both $\tilde{K}$ and $\tilde{K}'$ points; their common sign indicates that chirality of the pseudospinors at these points is the same. As a result, within this (blue) band (Fig.~\ref{fig-s3}b), we obtain a 
non-vanishing net Berry flux of $\mathcal{C} = -1$ in a single (original) valley 
(see Fig.~\ref{fig-s3}). 

The large Berry flux in the 
the top valence blue miniband (for $V_s<0$) arises from a very small minigap with the adjacent miniband just below it (green). Indeed, when zoomed-in a small minigap between the two top valence minibands of order $\sim \mu\, {\rm eV}$ appears at the $\tilde{K}'$ point (see Fig.~\ref{fig-s3}b); a similar gap also appears at $\tilde{K}$ point. We note parenthetically that this small gap is larger than the numerical mesh resolution from our numerical diagonalization routine, allowing us to extract the minigap value shown. Indeed, the lower miniband (green) also displays the same behavior with a sharp distribution of Berry curvature peaked at the $\tilde{K}, \tilde{K}'$ points; these have opposite sign to the blue miniband above it with a concomitant $\mathcal{C} = -1$ in a single valley.

While the finite valley Chern numbers indicate a non-trivial topology, the minigap is so small as to make it very difficult to meaningfully resolve any real experimental signatures of the finite valley Chern number. Indeed, for practical purposes, when energy resolution is larger than the small minigap $\sim \mu\, {\rm eV}$, the minigap cannot be resolved. As a result, the blue and green minibands close to $\tilde{K}, \tilde{K}'$ points resemble Dirac cones with the {\it same} chirality. This closely mirrors the behavior of the flat minibands in twisted bilayer graphene close to magic angle, where electrons close to $\tilde{K}, \tilde{K}'$ in the moir\'e Brilliouin zone are severely slowed, and possess the same chirality. The main difference between the green and blue bands in PDS scheme using TMDs is a reduced degeneracy (only two for spin).

\subsection{PDS induced minigaps}

In this section we discuss how the minigaps between the bundles of flatbands evolve as a function of applied superlattice potential $V_s$. Here we show the first minigap between minibands in the valence band (i.e., the gap between blue and green bands when $V_s > 0$, and the gap between green and purple band when $V_s < 0$, see Fig.~\ref{fig2} in the main text) as a function of $V_s$. This indicates that the minigaps increase as larger superlattice potential is applied. 

\begin{figure}[h!]
\includegraphics[width=\columnwidth]{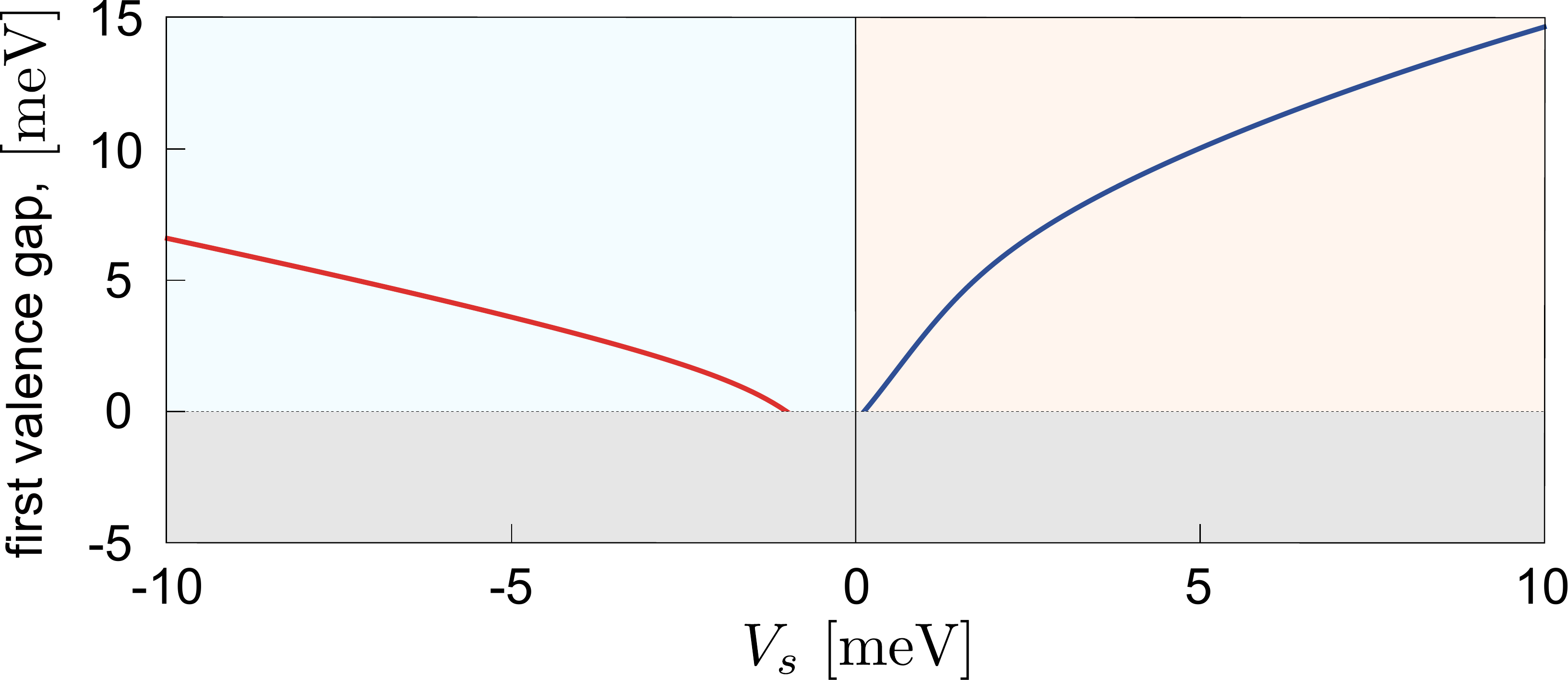}
\caption{PDS induced minigaps between blue and green minibands in the valence band (for $V_s>0$) as well as between green and purple minibands (for $V_s<0$). The gray shaded area indicates the region where the minibands are not well separated.}
\label{fig-s4}
\end{figure}

\subsection{Other types of superlattices}

\begin{figure*}[!]
\includegraphics[width=2\columnwidth]{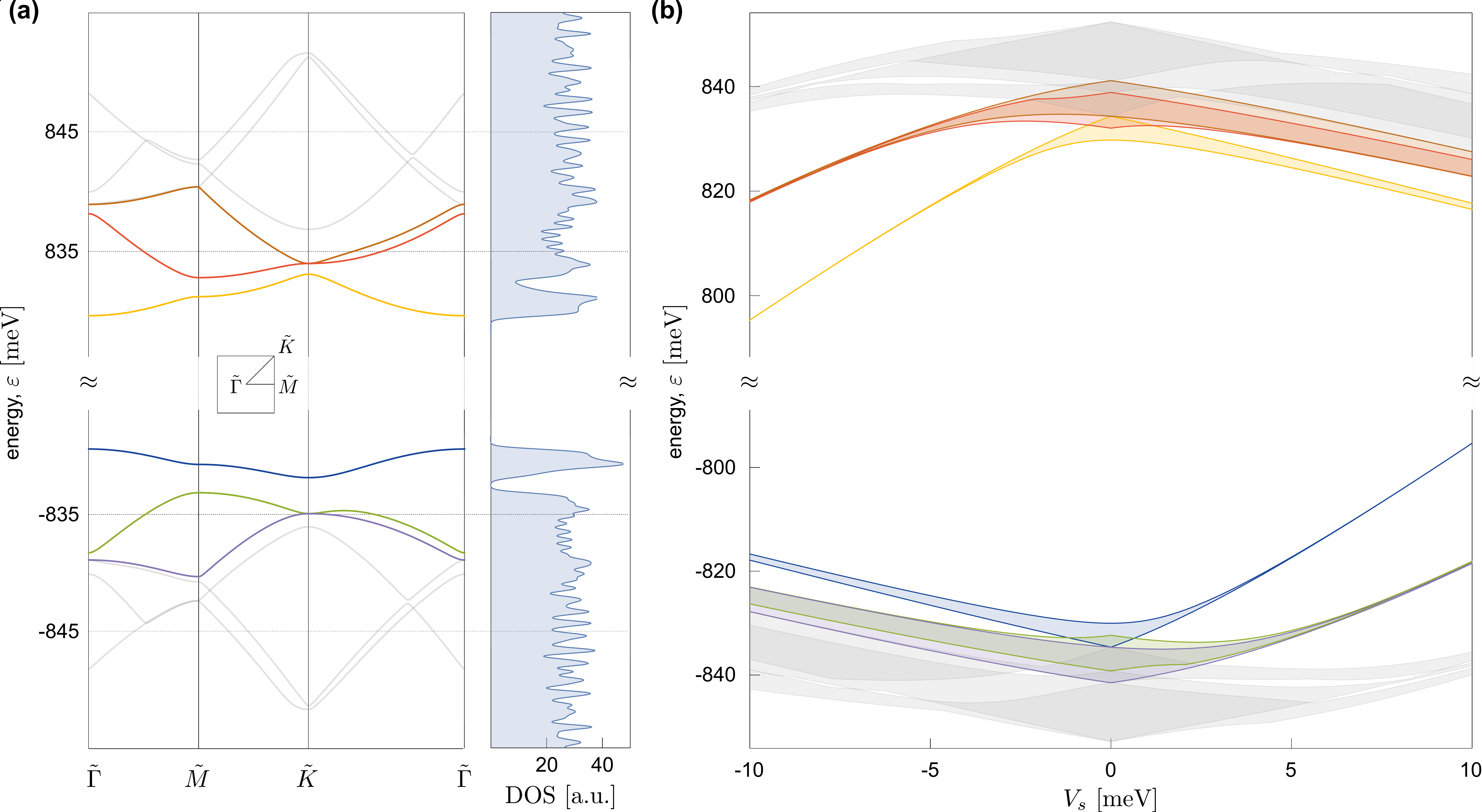}
\caption{Dispersion and DOS at $V_0 = 0.5~\text{meV}$ (a) and bandwidth (b) for a square superlattice with lattice spacing of 20 nm. Other parameters are the same with those in Fig.~\ref{fig2}a in the main text.}
\label{fig-s5}
\end{figure*}

\begin{figure*}[!]
\includegraphics[width=2\columnwidth]{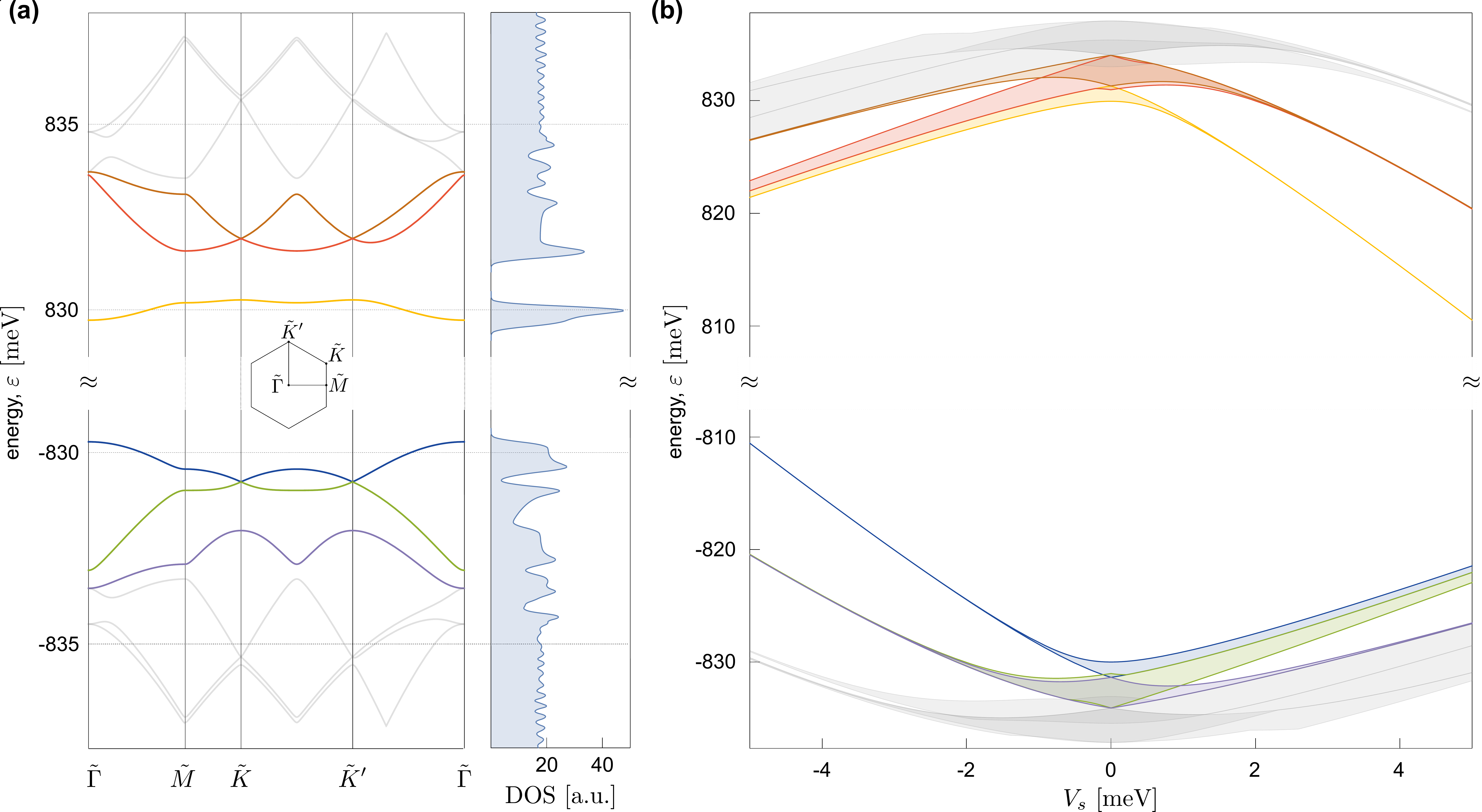}
\caption{Dispersion and DOS at $V_0 = 0.25~\text{meV}$ (a) and bandwidth (b) for a hexagonal superlattice with inter-hole spacing of 20 nm. Other parameters are the same with those in Fig.~\ref{fig2}a in the main text.}
\label{fig-s6}
\end{figure*}

Here we show that square and hexagonal superlattices can also lead to flat bands. For the square superlattice, we follow Ref.~\cite{Forsythe2018} and use the superlattice potential
\begin{equation}
V_\text{squ}(\vec r) = 2 V_s \sigma_0 [\cos (\vec b_1 \cdot \vec r) +\cos (\vec b_2 \cdot \vec r) + \cos (\vec b_1 \cdot \vec r) \cos (\vec b_2 \cdot \vec r) ] , 
\end{equation}
where $\vec b_1 = (2\pi/a) (1, 0)$ and $b_2 = (2\pi/a) (0, 1)$ with $a$ the superlattice spacing. For hexagonal superlattice, we use the superlattice potential
\begin{equation}
V_\text{hex} (\vec r) = V_s\sigma_0 \sum_{j=1,2,3} \big[\cos (\vec G_i \cdot \vec r) + \cos (\vec G_i \cdot \vec r')  \big], 
\end{equation}
where $\vec r' = \vec r + (a,0)$, and ${\vec G}_j $ are the same reciprocal vectors defined in Eq.~\ref{eq:potential} which also apply to hexagonal lattices. The reconstructed minibands are shown in Figs.~\ref{fig-s5} and \ref{fig-s6}, respectively. These display that flat minibands are also obtained for these superlattices underscoring the generic nature of the PDS flatband scheme. 

\subsection{Other types of TMDs}

\begin{figure}[h!]
\includegraphics[width=0.9\columnwidth]{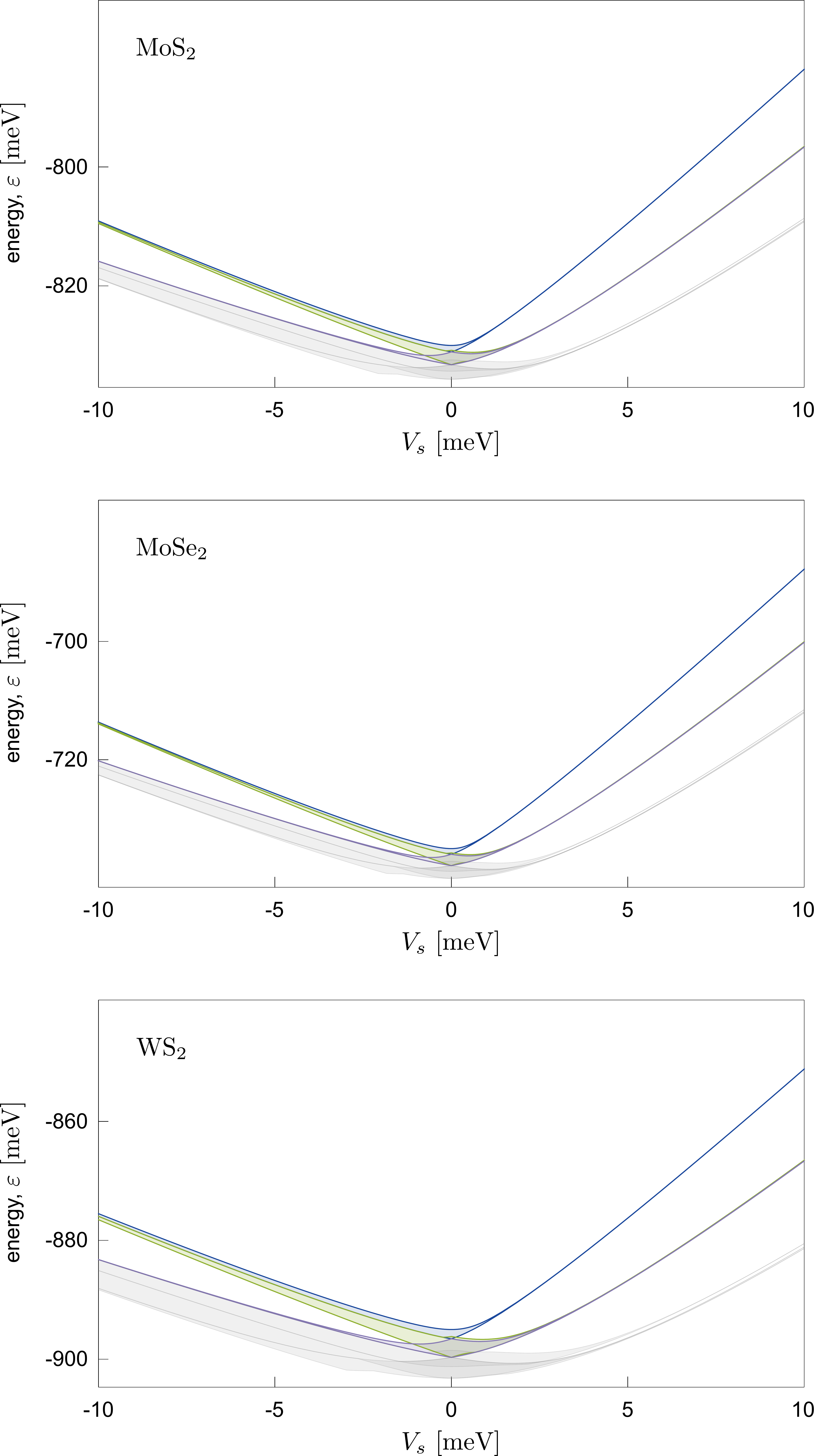}
\caption{
Valence miniband widths for triangular PDS for MoS$_2$ ($v \hbar = 3.51\,{\rm eV \AA}$, $\Delta = 0.88\,{\rm eV}$), MoSe$_2$ ($v \hbar = 3.11\,{\rm eV \AA}$, $\Delta = 0.73\,{\rm eV}$), and WS$_2$ ($v \hbar = 4.38\,{\rm eV \AA}$, $\Delta = 0.89\,{\rm eV}$). The lattice spacing $a = 20\,{\rm nm}$.}
\label{fig-s7}
\end{figure}

Here we discuss the PDS scheme (using triangular lattice as illustration) as applied to other types of TMDs, specifically, MoS$_2$, MoSe$_2$, and WS$_2$. The results are displayed in Fig.~\ref{fig-s7}. Here we only show the valence band widths, since the miniband structure possess a symmetry when $V_s \to - V_s$ and $\varepsilon_{\vec k} \to - \varepsilon_{\vec k}$ are interchanged (see the above section, ``Particle-hole asymmetry''). As expected, flat minibands can be readily achieved. 

\subsection{Estimate for disorder broadening $\Gamma$}

In this section, we estimate the disorder broadening. We begin by observing that 
charged impurity scattering is responsible for most of the observed bulk diffusive transport behaviors at low temperature (the contribution of phonons is frozen out). They reside either inside the substrate, or can be desposited near the interface between the substrate and the 2D system during the processing/sample preparation. Here we first extract the impurity concentration $n_\text{imp}$ from transport measurement results, and then use $n_\text{imp}$ to estimate the disorder broadening.
 
We will concentrate on modeling bare WSe$_2$, where recent low temperature transport measurements have revealed large mobilities $ \mu = 3 \times 10^4 ~{\rm cm^2 \, V^{-1} \, s^{-1} } $ at $T =1.4\,\text{K}$ for carrier density $ n=2 \times 10^{12}~\text{cm}^{-2} $~\cite{KimAPS}. To proceed, we first note that 
close to the band edge of a TMD, the dispersion can be written as 
$ \varepsilon_q = \sqrt{ v^2 \hbar^2  q^2 + \Delta^2}$
with $v \hbar = 3.94\,\text{eV}\,\text{\AA}$ and $\Delta = 0.8~\text{eV}$ \cite{Xiao2012}.
For small $q$, we adopt an effective mass approximation $\varepsilon_q \approx \Delta +  v^2 \hbar^2 q^2 / 2 \Delta $.
This produces
\begin{equation}
m^* = \Delta / v^2 = 0.39\,m_e ,
\label{eq:effmass}
\end{equation}
which is consistent with values obtained in the literature
~[S1,S2].
Also, within the effective mass approximation, its Fermi wave vector is
\begin{equation}
q_f = \sqrt{ \pi n } = 0.25 \, {\rm nm^{-1} } ,
\label{eq:qf}
\end{equation}
where we used $ n=2 \times 10^{12}~\text{cm}^{-2} $~\cite{KimAPS}. 

In order to extract an estimate for the impurity density from the mobility, we write 
the conductivity as 
\begin{equation}
\sigma = e n \mu = e^2 n \tau / m^* ,
\label{eq:conductivity}
\end{equation}
where $n$ is the carrier density (regardless of spin or valley), $\mu$ is the mobility,
and $\tau$ the scattering time.
At low temperature, the impurity scattering time reads as
\begin{equation}
\tau^{-1} 
= n_\text{imp}  \int \text{d} z \,  \la w (\phi_i ) (1-\cos \phi_i )  \ra_z ,
\label{eq:scattertime}
\end{equation}
where we assumed an uncorrelated charged impurities $n_\text{imp}$ (units ${\rm cm}^{-3}$) in the substrate, and $w$ is the scattering rate, see below. Eqs.~\ref{eq:conductivity} and \ref{eq:scattertime} lead to
\begin{equation}
n_\text{imp}
=   ( e / \mu m^* ) \Big[ \int \text{d} z \,  \la w (\phi_i ) (1-\cos \phi_i )  \ra_z \Big]^{-1} .
\label{eq:nimp}
\end{equation}
Within the effective mass approximation,
\begin{equation}
\la w ( \phi_i ) ( 1-\cos \phi_i )  \ra_z
= \int \frac{\text{d} \vec q}{4\pi \hbar} | V ( k , z) |^2
(1-\cos \theta ) 
\delta ( \varepsilon_q - \varepsilon_{q_i} )  ,
\end{equation}
\begin{equation}
V (q, z) =  \frac{2 \pi e^2}{ \epsilon (q + q_s) }  \exp (- q z ) ,
\label{eq:Vq}
\end{equation}
where $k = |\vec q - \vec q_i|$, $\theta = \phi_{\vec q} - \phi_{\vec q_i}$, $ \tan \phi_{\vec q} = q_y / q_x $, $\vec q_i$ is the initial electron wave vector at the Fermi surface with its magnitude equals to Fermi wave vector $q_f$, and $q_s$ is the effective screening wave vector.

Eqs.~\ref{eq:nimp} above enables to estimate $n_{\rm imp}$ from a known mobility. Using this impurity concentration, we can estimate the disorder potential width. To do so, we note that 
randomly distributed positive and negative charged impurities create fluctuations of the disorder potential. Taking uncorrelated impurity positions, 
the amplitude of the fluctuations \cite{Adam2007,Skinner2013,Skinner2014} is given by 
\begin{align}
\Gamma^2 = \la [\delta V (\vec r)]^2 \ra & = n_\text{imp}
\int \text{d} z \int \frac{ \text{d} \vec q}{(2\pi)^2} V^2 (q, z) .
\label{eq:bandwidthimpurity} 
\end{align}

\begin{figure}[t!] 
\includegraphics[width=1\columnwidth]{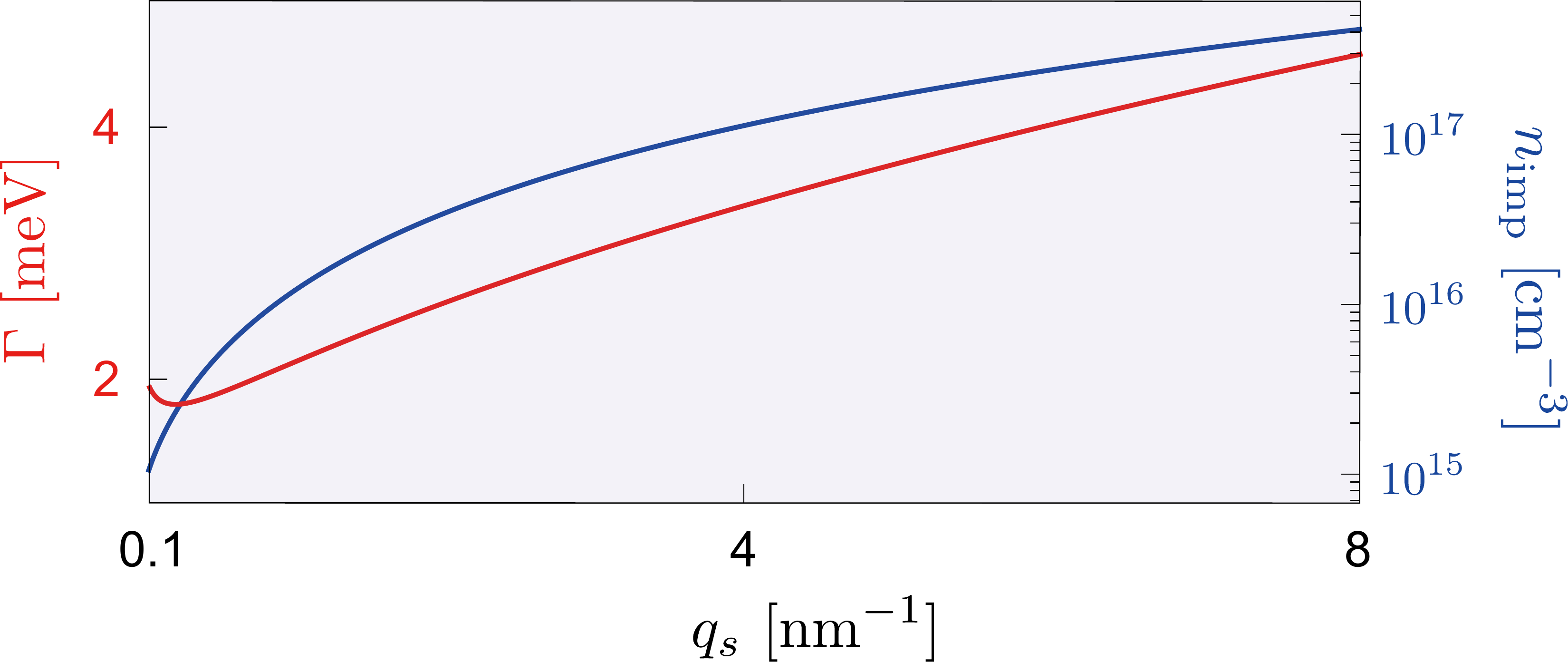}
\caption{Estimated disorder broadening $\Gamma$ (red) and impurity concentration $n_{\rm imp}$ (blue) for a wide range of effective $q_s$, for a WSe$_2$ sample having $n=2 \times 10^{12}~\text{cm}^{-2} $ and $ \mu = 3 \times 10^4 ~{\rm cm^2 \, V^{-1} \, s^{-1} } $ at $T =1.4\,\text{K}$ recently reported for WSe$_2$ from the Philip Kim group~\cite{KimAPS}.
}
\label{fig-s8}
\end{figure}

A numerical estimate for $n_{\rm imp}$ and $\Gamma$ both require the value for the screening wave vector $q_s = r_s^{-1}$. There are a number of processes that can contribute to screening. For example, at long-wavelengths, Thomas-Fermi screening~[S3] yields a screening wavevector as $ q_s = (2 \pi e^2/\epsilon) \nu_0 = (4 e^2 / \epsilon)(\Delta/v^2 \hbar^2) = 7.4 \,{\rm nm}^{-1} $,
where $\nu_0$ is the density of state at the Fermi energy, and $\epsilon = 4 $ for SiO$_2$. 
Screening can also arise from a proximal metallic top gate displaced several nanometers away from the vdW layer by $2 d$ \cite{Wu2018}, in this case $q_s \sim d^{-1}$ (assuming a perfect metallic gate). At low densities, non-linear screening can even take effect. The exact value of the disorder potential depends on details of how the Coulomb potential is screened. Instead of identifying the source screening which is beyond the scope of this work, here we take an {\it effective} screening wavevector approach. To obtain a rough estimate of the range of values $\Gamma$ can take on, we plot $\Gamma$ for a wide range of effective $q_s$ values (Fig.~\ref{fig-s8}). We find that for ultra-clean WSe$_2$ samples $\Gamma \sim$ several ${\rm meV}$ (see Fig.~\ref{fig-s8}). 

We note the above yields an estimate of disorder broadening for bare WSe$_2$. The precise disorder broadening  when the PDS scheme is applied will depend on the details of screening, as well as the actual impurity concentration of the PDS sample. The detailed form as well as the values for these are, at present, unavailable.  As a result, the disorder broadening obtained above represents an estimate for the possible disorder broadening in WSe$_2$ PDS flatbands. 

\subsection{Estimate for mean free path $\ell$}

Here we estimate the mean free path $\ell$ for the sample which has $ n=2 \times 10^{12} \, {\rm cm^{-2} } $, $q_f = \sqrt{ \pi n } \approx 0.25\,{\rm nm^{-1} }$ and $ \mu = 3 \times 10^4 \,{\rm cm^2 \, V^{-1} \, s^{-1} } $~\cite{KimAPS}:
\begin{equation}
\ell = v_f \tau = \frac{\hbar q_f}{ m^*}  \frac{\mu m^*}{e} = 0.49 \, {\rm \mu m}  .
\end{equation}
Since $\ell \gg a$, the electronic quasiparticles are able to experience Bragg scattering induced by the PDS superlattice potentials in this sample.

\subsection{Variational estimate for wavefunction extent $a_W$}
Here we use a variational approach to estimate $a_W$. We first adopt an effective mass approximation for an electron in conduction band:
\begin{equation}
\hat{H} = - \frac{\hbar^2  \nabla^2 }{ 2 m^*} + V(\vec r) ,
\end{equation}
where $V(\vec r) = - V_s \sum_{j=1,2,3}  2 {\rm cos} (\vec G_j \cdot \vec r) $ is the triangular superlattice potential used in the main text. The minus sign in the potential is to create a local potential minimum for an electron at $\vec r = \vec 0$. 

\begin{figure}[t!] 
\includegraphics[width=1\columnwidth]{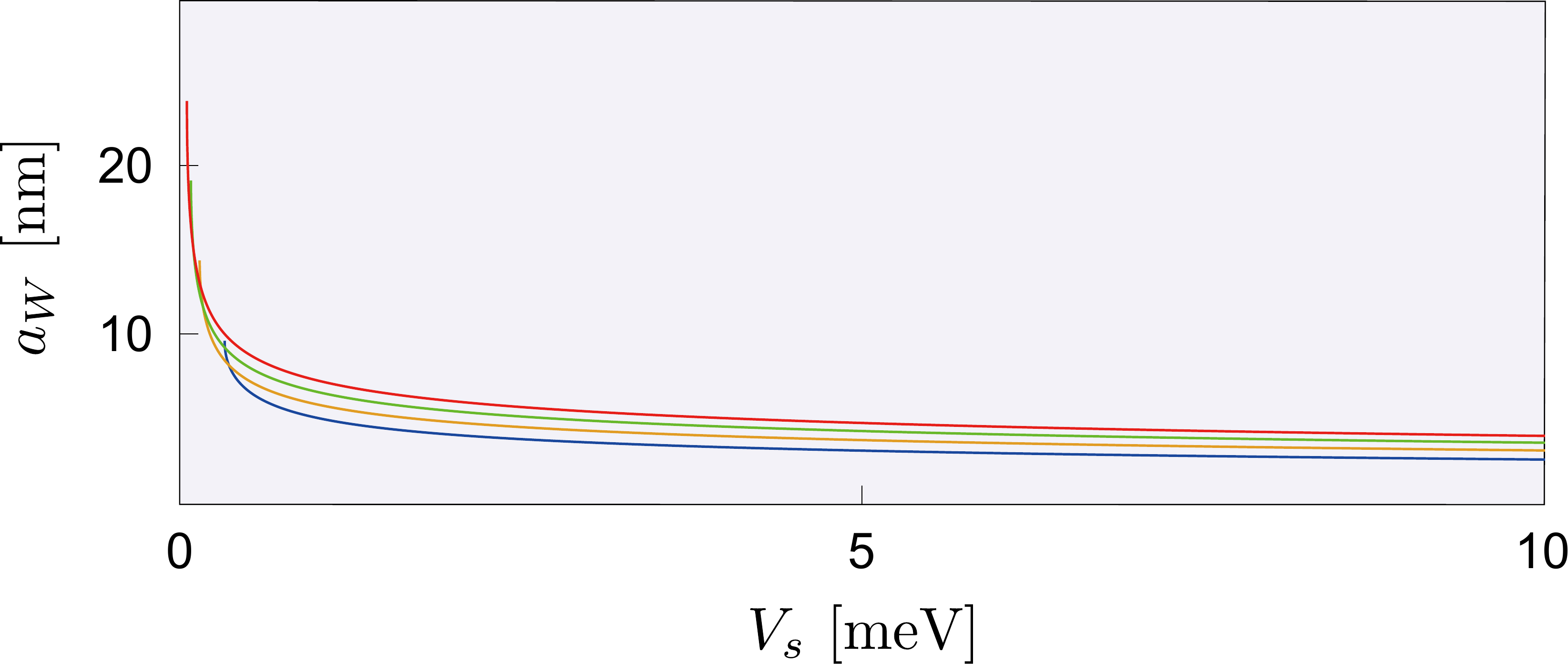}
\caption{Estimate of wavefunction spatial extent $a_W$ (as a function of applied $V_s$) when confined in a trough of a superlattice. This was obtained using a variational approach and with an effective mass approximation. Blue, yellow, green, and red lines correspond to superlattice wavelengths of $a =$ 20, 30, 40, 50 nm, respectively. For very small $V_s$, $a_w$ rapidly increases and gets cut off; below this critical $V_*$ the electrons are not well confined in the superlattice troughs. 
}
\label{fig-s9}
\end{figure}

To proceed in the variational approach, we use a Gaussian trial wave function
\begin{equation}
\psi (\vec r) = \frac{1}{ \sqrt{2 \pi a_W^2} } \exp \lp - \frac{\vec r^2}{ 4 a_W^2} \rp  ,
\quad
\int \text{d} \vec r [\psi (\vec r) ]^2 = 1 ,
\end{equation}
to describe an electron trapped at $\vec r = \vec 0$. The energy for this trial wave function is
\begin{equation}
\la \hat{H} \ra 
= \frac{\hbar^2 a_W^{-2} }{4 m^*} -  6 V_s \exp \lp - \frac{8 \pi^2}{9} \frac{ a_W^2 }{a^2}  \rp ,
\end{equation}
and the equation for its extremum reads as
\begin{equation}
\frac{\p \la \hat{H} \ra}{ \p a_W }  = 
\frac{32\pi^2}{3}\frac{a_W^2}{a^2 } V_s 
\exp \lp  - \frac{8 \pi^2}{9} \frac{ a_W^2 }{ a^2 } \rp
- \frac{\hbar^2 a_W^{-2} }{2 m }= 0.
\label{eq:variational}
\end{equation}
Eq.~\ref{eq:variational} has a finite and positive solution for $a_W$ that decreases with increasing $V_s$ ($\mathbf{e}$ is Euler's number):
\begin{equation}
a_W = \frac{ 3 a}{2 \pi} 
\big[ - W_0 \big( - \mathbf{e}^{-1} \sqrt{V_* / V_s } \big) \big]^{1/2} ,
~
V_* = \frac{ \mathbf{e}^2 \pi^2 }{ 54 } \frac{ \hbar^2 a^{-2} }{ 2 m^* } .
\end{equation}
Here $W_0 (z) $ is the the Lambert $W$ function,
and $W_0 (z) $ only has a real, negative solution when $- \mathbf{e}^{-1} \le z \le 0 $. Specifically, $ W_0 (- \mathbf{e}^{-1}) = - 1$ and $ W_0 (0) = 0$.
Therefore, $V_s \ge V_*$ is required for consistency of our initial bound state trial wave function/ansatz, i.e., a finite and positive $a_W$. When $V_s < V_*  $, $a_W$ becomes a complex number, indicating the bound state trial wave function/ansatz fails. As a result, only a large enough $V_s \ge V_*$ is able to trap an electron at its local minimum, and to create a nearly flat band. Thankfully for the TMDs we consider in this work, this critical $V_*$ is small due to their large effective masses (see Fig.~\ref{fig-s9}). To illustrate the typical extent of the trapped wavefunction we plot the behavior of $a_W$ versus $V_s$ in Fig.~\ref{fig-s9}. Here we have used $m^* = 0.39\, m_e$ with $a=20,30,40, 50 \, {\rm nm}$ corresponding to blue, yellow, green and red lines, respectively. As displayed, even fairly modest values of $V_s$ yield fairly localized $a_W$. 

\bigskip

\subsection{Flat band ferromagnetism}

After discussing how nearly flat bands can be created in TMDs using the PDS scheme, 
we now demonstrate a possible spontaneous symmetry breaking that arises due the quenching of kinetic energy.

We start from the electronic Hamiltonian 
\begin{equation}
H = \sum_{\vec k \sigma} 
\xi_{\vec k \sigma} c_{\vec k \sigma}^\dagger c_{\vec k \sigma}^{} + H_\text{int} ,
\end{equation}
where $\xi_{\vec k \sigma}$ describes the dispersion, and
\begin{equation}
H_\text{int} = \frac{U}{2 {\cal V}} 
\sum_{\vec k' \vec k \vec q, \sigma \sigma'}  
c_{\vec k + \vec q \sigma}^\dagger 
c_{\vec k' - \vec q \sigma'}^\dagger
c_{\vec k' \sigma'}^{}
c_{\vec k \sigma}^{} ,
\end{equation}
is the electron-electron interaction. Here we only retain the on-site repulsive interaction $U$ and neglect inter-site interactions, which are strongly suppressed in nearly flat bands \cite{Wu2018}. 
Generally, spin polarization in the $z$-direction is favored due to repulsive interactions~[S4]
thus we will focus on the possibility of the spin polarization in the $z$-direction, and use 
the mean field parameterization as
\begin{equation}
\la c_{\vec k \sigma}^\dagger c_{\vec k' \sigma'}^{} \ra 
= \delta_{\vec k \vec k'} \delta_{\sigma \sigma'} \bar{n}_{\vec k \sigma} .
\end{equation}

Using this meanfield ansatz, we obtain the 
mean field Hamiltonian
\begin{equation}
H_\text{MF} = \sum_{\vec k \sigma} \lp \xi_{\vec k} + U \bar{n}_{\bar{\sigma}} \rp c_{\vec k \sigma}^{\dagger} c_{ \vec k \sigma}^{} 
- U {\cal V} \, \bar{n}_{\uparrow}  \bar{n}_{\downarrow} ,
\end{equation}
where $\bar{\sigma}$ is opposite to $\sigma$,
and $ \bar{n}_\sigma = {\cal V}^{-1} \sum_{\vec k} \la c_{\vec k \sigma}^{\dagger} c_{ \vec k \sigma}^{} \ra_\text{MF} $ is the spin $\sigma$ density with respect to the mean field ground state.
For a nearly flat band, we can neglect the $\vec k$-dependence of $\xi_{\vec k} \equiv \xi_0$, and its energy density is
\begin{equation}
\la \bar{H}_\text{MF} \ra =
\lp \xi_0 + U \bar{n}_{\downarrow} \rp \bar{n}_{\uparrow} 
+
\lp \xi_0 + U \bar{n}_{\uparrow} \rp \bar{n}_{\downarrow}
- U \, \bar{n}_{\uparrow}  \bar{n}_{\downarrow}  ,
\end{equation}
where $\la \bar{H}_\text{MF} \ra = \la H_\text{MF} \ra / {\cal V} $. In anticipation of the (spin-split) broken symmetry state, we can re-write $\la \bar{H}_\text{MF} \ra$ to emphasize that the exchange interaction favors aligned spins:
\begin{equation}
\la \bar{H}_\text{MF} \ra = \sum_\sigma \lp \varepsilon_0 \bar{n}_\sigma - \frac{U}{2} \bar{n}_\sigma^2 \rp ,
\label{eq:meanfieldflatband}
\end{equation}
where $\varepsilon_0 = \xi_0 + U ( \bar{n}_\uparrow + \bar{n}_\downarrow )$.

When there is a finite bandwidth $\Gamma$ [for e.g., arising from disorder broadening (see main text and above)], $\varepsilon_0$ no longer resides at a single flat energy, but instead fluctuates. To capture this, the [single-particle] energy density can be described by a broadened spectral function. Here we have used a simple spectral function $Z (\varepsilon) = A \exp [ - (\varepsilon - \varepsilon_0)^2 / 2\Gamma^2 ]  $ to model this level broadening, where $A = [ \sqrt{2\pi \Gamma^2} ]^{-1}$. Other choices of spectral function do not modify the qualitative conclusions we describe below and in the main text. As a result, the energy density can be described in close analogy with that used for the energy density in Landau levels \cite{Nomura2006}.

Using the spectral function above, we find the energy density (including disorder broadening) can be modeled as 
\begin{equation}
\la \bar{H}_\text{MF} \ra =
\sum_\sigma \lb \int^{\mu_\sigma} \text{d} \varepsilon \, \varepsilon Z (\varepsilon) - \frac{U}{2} \bar{n}_\sigma^2 \rb ,
\label{eq:meanfieldbroadened}
\end{equation}
where $\mu_\sigma$ is the Fermi level for spinor-component $\sigma$ that satisfies
\begin{equation}
\bar{n}_\sigma = \int^{\mu_\sigma} \text{d} \varepsilon \, Z(\varepsilon) .
\end{equation}
As we can see, Eq.~\ref{eq:meanfieldbroadened} mirrors Eq.~\ref{eq:meanfieldflatband} except that the single-particle energy density is now broadened by the spectral function (due to disorder broadening). 

In the normal state the two degenerate flat bands are equally occupied. To see whether it is a stable state, we perturb $\bar{n}_\sigma = \bar{n}_0 + \delta \bar{n}_\sigma$ with $\bar{n}_0 = ( \bar{n}_\uparrow + \bar{n}_\downarrow ) / 2$. By using $\sum_\sigma \delta \bar{n}_\sigma = 0$ and expanding $\delta \mu_\sigma$ to the second order, the meanfield energy density reads as
\begin{align}
\la \bar{H}_\text{MF} \ra
=~& \sum_\sigma \lb \int^{\mu_0} \text{d} \varepsilon \, \varepsilon Z (\varepsilon) - \frac{U}{2} \bar{n}_0^2 \rb
\nn
+ & \sum_\sigma \frac{\delta \bar{n}_0^2}{2} \lb \frac{1}{Z (\mu_0)} - U \rb + \cdots .
\end{align}
From this we see the normal spin-degenerate state is unstable when the second term is negative, i.e., $ Z (\mu_0) U > 1$.

\clearpage
\end{document}